\documentclass[hyper]{JHEP3} 

\usepackage{amsmath}
\usepackage{graphicx}
\usepackage{epsfig}
\usepackage{cite}

\makeatletter

\newcommand{\gsim}{\lower.7ex\hbox{$\;\stackrel{\textstyle>}{\sim}\;$}}
\newcommand{\lsim}{\lower.7ex\hbox{$\;\stackrel{\textstyle<}{\sim}\;$}}
\newcommand{\gev}{\,{\rm GeV}}

\def\beq{\begin{equation}}
\def\eeq{\end{equation}}
\def\bea{\begin{eqnarray}}
\def\eea{\end{eqnarray}}

\title{Bounds on the Fermion-Bulk Masses in Models with Universal Extra Dimensions}

\author{Gui-Yu Huang, Kyoungchul Kong\\
        Department of Physics and Astronomy, University of Kansas, Lawrence, KS 66045 USA \\
        E-mail: \email{huang@ku.edu, kckong@ku.edu}
        }

\author{Seong Chan Park\\
	Department of Physics, Sungkyunkwan University, Suwon 440-746, Korea\\
	E-mail: \email{s.park@skku.edu}
        }

\abstract{
In models with extra dimensions,  vectorlike Dirac masses for fermion fields are generically allowed. 
These  masses are independent of electroweak symmetry breaking and do not contribute to the known masses for the quarks and leptons.
They control the profile of the bulk wave functions, the mass spectra of Kaluza-Klein modes, and interactions that could be tested in experiments. 
In this article, we study the effects of bulk masses in electroweak precision measurements and in dark matter and collider searches, 
to set  bounds on the bulk mass parameters in models with a flat universal extra dimension, namely,  Split-UED. 
We find the current bound on the universal bulk-mass to be smaller than (0.2-0.3)/$R$, where $R$ is the radius of the extra dimension. 
Similar but slightly relaxed bounds are obtained in the non-universal bulk mass case.
The LHC is expected to play an important role in constraining the remaining parameter space.
}


\begin{document}

\section{Introduction}
\label{sec:intro}

As one of the most attractive extensions of the standard model (SM),  extra
dimensions have been extensively considered to address various problems in
particle physics and cosmology. In particular, models
with Universal Extra Dimensions (UED) \cite{Appelquist:2000nn} not only offer
interesting dark matter phenomenology,  but also predict signals that
can be tested at colliders.  Among many different scenarios, a 5-dimensional
version, often referred to as Minimal Universal Extra Dimensions (MUED)
\cite{CMS,Cheng:2002ab}, has been studied in detail (for reviews of the UED model
and its phenomenology, see Ref.~\cite{Hooper:2007qk,Datta:2010us,Kong:2010mh}). 
Recent studies on MUED suggest a lower bound of  $700 \gev$ \cite{collbounds} on the KK mass scale $1/R$ (where $R$ is the radius of the extra dimension) 
from the first year LHC data \cite{Arrenberg:2008wy,Bertone:2010ww}. 
One also expects a relatively weak lower bound of $1/R  \gsim 600 \gev$  from flavor constraints \cite{flavorbounds}, and 
$1/R \gsim 750 \, (300) \, \gev$ for $m_h=115\, (750) \,\gev$ \cite{ewAY,ewGM,ewGfitter} from electroweak (EW) precision
measurements mainly due to the presence of (approximate) Kaluza-Klein (KK) number 
conservation and KK parity conservation. KK parity guarantees the stability of the lightest KK particle (LKP), thus providing a viable dark matter candidate
\cite{servanttait,Cheng:2002ej}.  In general, computation of relic abundance leads
to an upper bound on the mass scale \cite{servanttait,Kong:2005hn,Belanger:2010yx,Kakizaki:2006dz,Kakizaki:2005uy,Kakizaki:2005en}, resulting in a tension with electroweak constraints.   

This tension may be relieved by considering effective coannihilation processes involving level-2 KK-leptons.
A recent study \cite{Belanger:2010yx} shows that the preferred dark matter scale in MUED could be increased to $1/R\simeq 1.4$ TeV, 
even though a high level of degeneracy, within a few percent, between the dark matter particle 
(level-1 KK state of $U(1)_Y$ gauge boson, $B_1$ or KK photon $\gamma_1$) and level-1 KK-leptons  ($\ell_1$) has to be realized.  
The high level degeneracy is a consequence of an assumption that brane localized operators at orbifold fixed points \cite{BLKTrefs} 
as well as 5D fermion mass terms \cite{sUED1} are all vanishingly small, which leads us to another fine-tuning problem. 
Indeed, the dark matter and collider phenomenologies of UED models strongly depend on
the detailed KK mass spectrum, which could be modified when either operators at the
orbifold fixed points or 5D fermion mass terms  are taken into account.  In either
case, not only the KK mass spectrum but also the KK particle couplings are significantly
modified. In particular, the couplings of zero-mode fermions to even-numbered KK
gauge bosons  are generally non-zero, even at tree-level.  Such couplings imply
that higher KK modes can appear as resonances at the LHC \cite{2ndKKrefs,sUED2,sUEDWp}.

Recently electroweak constraints on 5D fermion masses have been examined in Ref. \cite{arXiv:1111.7250} in a concrete context of Split Universal Extra Dimension (SUED) model \cite{sUED1}.
For a large bulk mass ($\mu R \gsim 1$, where $\mu$ is the size of the bulk mass), 
the lower limit on the allowed KK mass scale, $R^{-1}$, is stronger than
that in MUED, while close to the MUED limit ($\mu R\lesssim 1$), we notice that
the corresponding lower bounds are relaxed allowing for lower KK mass scales.
However it is important to compare with the preferred mass scale given by cosmological observations. 
A naive expectation is that once KK fermion masses become heavier than those in MUED, 
the annihilation cross sections of KK photon get suppressed by masses of the mediating KK fermions in the  $t$- and $u$-channels, 
resulting in larger abundance hence lower KK scale. 
Therefore relic abundance requirement tends to lower the KK mass scale in the presence of fermion-bulk masses.

In this article, we consider constraints on fermion-bulk masses in SUED from various sources, such as relic abundance of dark
matter, collider searches and electroweak precision measurements. 
We find that a tension between electroweak precision
constraints and relic abundance still exists even in the presence of universal fermion-bulk mass. 
The tension may be weakened by introducing multiple fermion-bulk masses. 
It would be natural to have  two different masses in hadronic and leptonic sectors, separately.  Oblique corrections are mostly sensitive to KK tops due to a large Yukawa coupling, while the lepton sector may affect EW precision observables through the Fermi constant. Also the dominant contribution of relic abundance of KK photon dark matter comes from the lepton sector due to the nature of hypercharge interaction of the LKP. 
Finally, important collider limits in the presence of bulk masses arise 
differently from dijet and dilepton searches. 
For all these reasons, we consider a non-universal bulk mass case with two parameters. 
We consider a five dimensional version of SUED in this paper 
but expect that some of our results are still valid in different models with flat extra dimensions such as 6D UED, 
although additional constraints may arise. 

We begin with a brief review on SUED in section \ref{sec:sued}, 
followed by discussions on various constraints for the universal case in section \ref{sec:universal} 
and for the non-universal case in section \ref{sec:nonuniversal}.  
In each section we consider oblique corrections, relic abundance of KK photon, four Fermi interactions, 
anomalous muon magnetic moment and collider bounds.  
Section \ref{sec:summary} is reserved for conclusions.

\section{Universal Extra Dimensions with bulk fermion masses: split-UED}
\label{sec:sued}

%
%
Here we provide a brief review on  (Split) UED model with the exact KK-parity \cite{sUED1} following Ref.\cite{Csaki:2010az}, where the model is understood as a low energy effective description of Randall-Sundrum (RS) model.  The basic observation is that a
$\mathbb{Z}_2$ symmetric space can be constructed by gluing two identical spaces
together. The combined space is thus invariant under reflection about
the midpoint.
 We can identify this $\mathbb{Z}_2$ reflection symmetry as
the origin of KK parity.  
All SM particles are promoted to five dimensional
fields propagating in this ${\mathbb Z}_2$ symmetric space. 

\begin{figure}[t]
\centering
\epsfig{file=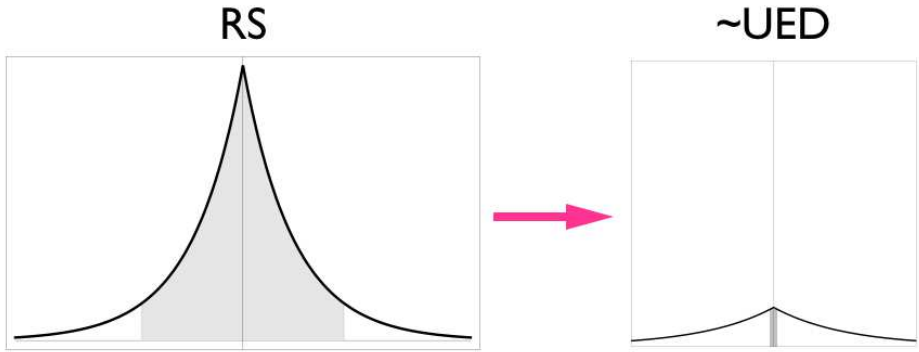, width=0.8\columnwidth}
\caption{\label{fig:doublethroat} \sl Schematic picture from Ref. \cite{Csaki:2010az} showing 
how integrating out parts of RS space yields  an effective ``UED'' space.}
\label{fig:RStoUED}
\end{figure}

As an explicit example, we glue two warped `throats' together. 
After integrating out the middle portion corresponding to the highly warped UV-regime,
 as illustrated in Fig.~\ref{fig:RStoUED},
we obtain an effective geometry largely determined 
by the vicinity of the IR-boundaries, the near-flatness of which is
desired for UED models.
 The integrating-out preserves the $\mathbb{Z}_2$ symmetry.
Some phenomenological features of the two warped throat model were studied
earlier in Refs.~\cite{Cacciapaglia:2005pa, Agashe:2007jb}. 

In Split UED, Dirac masses for the fermions in the bulk are introduced 
 for all fermions $\Psi=(Q,U,D,L,E)$ 
\beq
S_{\rm SUED}\ni -\int d^4x\int_{-L}^{L} d \, y M_\Psi (y) \overline{\Psi}\Psi \, ,
\eeq
where the `size' of the extra dimension is denoted by $L=\frac{\pi R}{2}$, and $y$ is the coordinate of the extra dimension. 
With such a mass term, the fermion KK spectra can be significantly modified 
and {\it split} from those of the KK gauge bosons.  To preserve KK parity, the
$y$-dependent mass  must be odd under reflection $M_\Psi(y)=-M_\Psi(-y)$
\cite{sUED1}.  One notices that the mass term is compatible with {\it all}
required symmetries of the model, namely, the 5D Lorentz symmetry, KK
parity and also the gauge symmetries ($SU(3)_C \times SU(2)_W\times U(1)_Y$
 in the minimal setup), so it is natural to be included.  We choose the
simplest kink-type mass, which may arise from the minimum energy 
configuration with $w(y)\propto \tanh(m y)$ in the large mass limit: $m\to
\infty$,  where $w(y)$ is the profile of the scalar field of mass $m$ which is responsible for the fermion-bulk mass \cite{Georgi:2000wb}
\begin{eqnarray}
M_\Psi(y)&&=\mu_\Psi \tanh m y ~~~
\xrightarrow[]{m\to \infty}  ~~~ \mu_\Psi  \theta(y)
=\begin{cases}
-\mu_\Psi & \text{ if } y<0 \\ 
 +\mu_\Psi & \text{ if } y>0 
\end{cases}.
\end{eqnarray}
In order to avoid the restrictive flavor bounds, 
we will assume one 5D mass for all quarks 
 $-M_Q=M_U=M_D=\mu_Q\, \theta(y)$ and one for all leptons $-M_L=M_E=\mu_L\,\theta(y)$ 
 following Refs. \cite{sUED1, sUED2}, 
where in the universal bulk mass limit $\mu_L = \mu_Q\equiv \mu$. 

Here we summarize some results from Refs. \cite{sUED1, sUED2}, which are most
relevant for our phenomenological study (for details, see the original
references).  The zero-mode fermions remain massless before electroweak symmetry
breaking and obtain masses through Yukawa interactions. 
The mass of $n$-th KK level fermion follows from 
 \begin{eqnarray}
 m^2_{\Psi^{(n)}}= \begin{cases}
\lambda^2_\Psi v^2 & \text{ if } n=0 \\ 
\mu^2 + k^2_n+ \lambda^2_\Psi v^2 & \text{ if } n\geq 1 
\end{cases},
 \label{eq:mf}
 \end{eqnarray}
where $\lambda_\Psi$ is a Yukawa coupling and $v\approx 246\, {\rm GeV}$ is the
vacuum expectation value of the Higgs boson.  The `momentum' $k_n$ is determined as
 \begin{eqnarray}
 \label{eq:kn}
 k_n =\begin{cases}
\begin{cases}
i\kappa_1 & :\kappa_1=\kappa \in\left\{0<\kappa  ~ | ~\mu=-\kappa \coth \kappa L, \,\mu L<-1\right\} \\ 
k_1 & :k_1=k \in \left\{0 \leq k\leq \frac{\pi}{L} ~| ~\mu=-k \cot k L,\,\mu L\geq -1\right\} 
\end{cases} & \text{ for } n=1 \\ 
 \frac{n}{R}= \frac{n\pi}{2 L}& \text{ for } n=2,4,6,\cdots \\ 
k_n =k \in \left\{\frac{(n-2)\pi}{L}<k<\frac{(n-1) \pi}{L}  ~ | ~  \mu=-k \cot k L \right\} &\text{ for } n=3,5,7,\cdots \, .
\end{cases}
\end{eqnarray}

\begin{figure}[t]
\centerline{
\epsfig{file=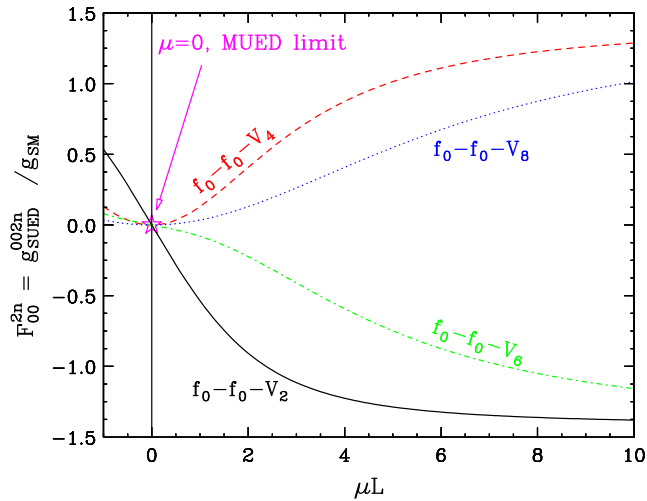, width=.55\textwidth}  }
\caption{\sl The ratio of tree-level couplings in SUED to the corresponding SM couplings.}
\label{fig:002n}
\end{figure}

The coupling constants between the $n^{\rm th}$ KK gauge boson and $m^{\rm th}$
and $\ell^{\rm th}$ KK fermions are determined by the overlap integral of the
wave functions, 
\begin{eqnarray}
g_{m\ell n} &=& \frac{g_5}{\sqrt{2L}}\int_{-L}^L dy \, \psi_m(y)\psi^*_\ell(y) f^n_V(y) \\
                    &\equiv&  g_{\rm SM}{\cal F}_{m\ell}^n(\mu_\Psi L)    \,  ,
\end{eqnarray}
where $g_{\rm SM}\equiv g_5/\sqrt{2L}$ is identified with the respective SM couplings and $\psi_{n}$ ($f_V^n$) is the wave function
of the  $n^{\rm th}$ KK excitation of the fermion (vector boson).
One immediately verifies that ${\cal F}_{00}^0=1$ irrespective of $\mu$, as expected. 
We  also emphasize that no KK number violating couplings exits between zero-mode gauge bosons 
and KK fermions ($V_0$-$f_n$-$f_{\ell}$) for $n\neq \ell$ as ${\cal F}_{n \ell}^0=0$.  
The zero-mode fermions do not couple to
gauge bosons at odd KK levels, respecting KK parity.  On the other hand, the 
coupling of the $(2n)^{\rm th}$ gauge bosons to the SM fermion pair is allowed as \cite{sUED2} 
\begin{eqnarray}
g_{002n}&=&g_{\rm SM} {\cal F}^{2n}_{00}(x_\Psi=\mu_\Psi L) \label{eq:g002n1}\\
                &=& g_{\rm SM} \frac{x_\Psi^2 \left[1-(-1)^n e^{2x_\Psi}\right]\left[1-\coth\left(x_\Psi \right)\right]}{\sqrt{2(1+\delta_{0n})}\left[x_\Psi^2+n^2 \pi^2/4 \right]} \, . \label{eq:g002n2}
\label{eq:coupl}
\end{eqnarray}
A few examples of the $(2n, 0, 0)$ couplings are shown as functions of the bulk mass term
in Fig. \ref{fig:002n}. 
The star (in magenta) represents the MUED limit ($\mu \to 0$) where those couplings vanish due to KK number conservation in MUED at tree level (later we mention loop-induced KK number violation). 
For $\mu \to \infty$, the zero mode fermions are well localized near the center (y = 0) so that their couplings to KK gauge bosons asymptotically approach  the well known value $(-1)^{n} \sqrt{2}$ as one can see in Fig. \ref{fig:002n}. 
The alternating sign arises from the $2n$-th KK gauge boson wave function which is proportional to $\cos n\pi = (-1)^n$ at y = 0 
where the fermion wave function is mostly localized, while the $\sqrt{2}$ is from the zero mode normalization. 

Many phenomenological aspects of SUED
model have been explored, including dark matter \cite{Chen:2009gz}, collider
\cite{Chen:2009gz,sUED2, SUED3, Edelhauser:2010gb}, electroweak constraints \cite{arXiv:1111.7250} and flavor structure \cite{Csaki:2010az}.  For studies on other varieties of UED models, see  Refs. \cite{pUED, 6DUED}.

In the following two sections we examine various constraints on the parameter space of Split UED.
The two cases we study are one with a universal bulk mass, and the other with simplified non-universal
bulk mass terms $\mu_Q$ and $\mu_L$.
We consider oblique corrections in terms of $(S,T,U)$ parameters,  the
relic abundance of the KK dark matter, four Fermi contact interactions,
anomalous magnetic moment of muon and also collider bound mainly from the recent LHC data.

\section{The Universal  Case: $\mu_L=\mu_Q$} 
\label{sec:universal}

\subsection{Oblique Corrections}
\label{sec:stu}

The constraints on the MUED model from electroweak precision tests have
been studied in Refs.~\cite{ewAY,ewGM,ewGfitter} by calculating the MUED
contributions  $(S, T, U)_{\rm UED}$ to the oblique parameters
\cite{PT1} at one loop-level. The leading contributions are found in 
Refs.~\cite{ewAY,ewGM,ewGfitter}:
\bea
S_{UED}&=&\frac{4 \sin^2\theta_W}{\alpha}\left[\frac{3g^2}{4(4\pi)^2}\left(\frac{2}{9}\sum_n \frac{m^2_t}{(n/R)^2}\right)+\frac{g^2}{4(4\pi)^2}\left(\frac{1}{6}\frac{m^2_h}{1/R}\right)\zeta(2)\right], \label{Sloop}\\
T_{UED}&=&\frac{1}{\alpha}\left[\frac{3g^2}{2(4\pi)^2}\frac{m_t^2}{m_W^2}\left(\frac{2}{3}\sum_n \frac{m^2_t}{(n/R)^2}\right)+\frac{g^2\sin^2\theta_W}{(4\pi)^2\cos^2\theta_W}\left(-\frac{5}{12}\frac{m^2_h}{1/R}\right)\zeta(2)\right], \label{Tloop}\\
U_{UED}&=&-\frac{4\sin^2\theta_W}{\alpha}\left[\frac{g^2\sin^2\theta_W}{(4\pi)^2}\frac{m_W^2}{(1/R)^2}\left(\frac{1}{6}\zeta(2)-\frac{1}{15}\frac{m_h^2}{(1/R)^2}\zeta(4)\right)\right] \, , \label{Uloop}
\eea
where $\theta_W$ is the Weinberg angle, $g$ is the coupling strength of $SU(2)_W$ interaction, and $\alpha=e^2/4\pi$ is the fine structure constant, and  
$m_t$, $m_W$ and $m_h$ are masses of the top, $W$ and Higgs in the SM, respectively.
$\zeta(n)$ is the Riemann zeta function.  
We neglect the small radiative corrections to KK masses. 
The terms proportional to the Riemann zeta functions arise from summations over
all KK Higgs and KK gauge boson loops.
Summation over KK tops is shown explicitly above for further discussion. 
All other KK fermion loops are neglected due to fermion mass suppression. 
For $m_h =120$ GeV, oblique corrections in MUED lead to $R^{-1} \gsim 700$ GeV at 95\% Confidence Level (C.L.). 

In Split UED, there appear additional corrections to the parameters via
KK $W$ contributions to the Fermi constant \cite{arXiv:1111.7250},
\begin{eqnarray}
G_F &=& G^0_F+\delta G_F  \, , \\
G^0_F &=& \frac{g^2}{\sqrt{32}  m_W^2} \, ,\\
\delta G_F &=& \frac{1}{\sqrt{32}}\sum_n\frac{g^2_{002n}}{m_W^2+\left(\frac{2n}{R}\right)^2} \, , \label{eq:deltaGF}
\end{eqnarray}
where $g_{002n}$ is defined earlier in Eq. (\ref{eq:coupl}). 
Treating the leptonic and hadronic sectors universally, $\mu=\mu_L=\mu_Q$ as in Ref. \cite{arXiv:1111.7250}, 
one can express the oblique parameters as follows \cite{CTWP}:
\bea
S_{SUED}&=&S_{UED},\nonumber\\
T_{SUED}&=&T_{UED}-\frac{1}{\alpha}\frac{\delta G_F}{G_F} ,\nonumber\\
U_{SUED}&=&U_{UED}+\frac{4 \sin^2\theta_W}{\alpha}\frac{\delta G_F}{G_F}.
\label{STUeff}
\eea 
Here the following substitution of the fermion
KK tower summations in the $(S,T,U)_{\rm UED}$ is understood:
\beq
\sum_n \frac{m^2_f}{(n/R)^2} ~ ~\to ~~ \sum_n \frac{m^2_f}{\mu^2 +k_n^2 + m_f^2} \, ,
\label{eq:replace}
\eeq
where $k_n$ is defined in Eq. (\ref{eq:kn}). 

\begin{figure}[t]
\centerline{
\epsfig{file=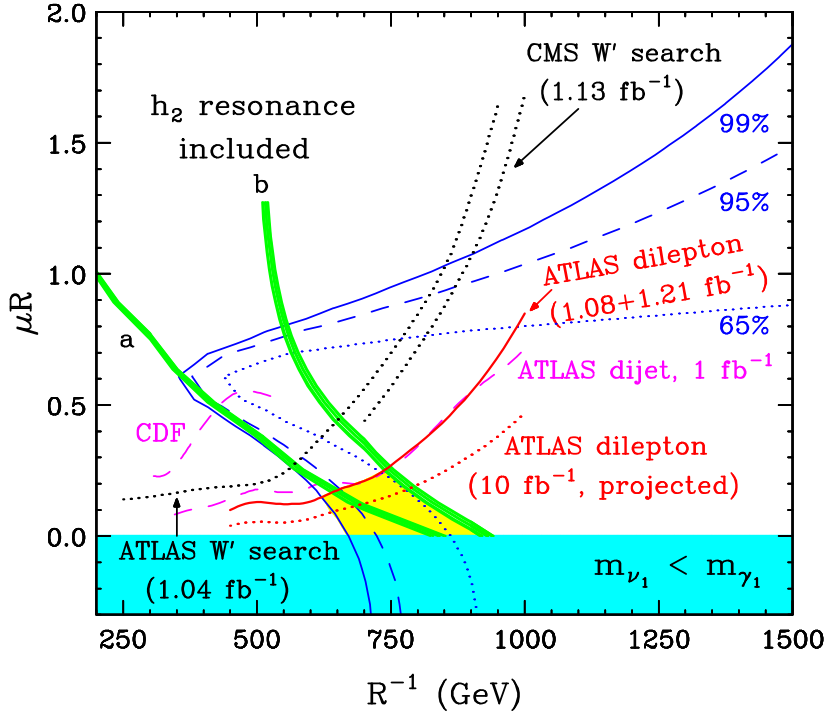, width=.7\textwidth}  }
\caption{\sl Constraints from relic abundance (in green, $\Omega h^2 = 0.1123 \pm 0.0035$), 
electroweak precision measurements (in blue) and collider search (in black, red, and magenta) 
in the Split UED parameter space for the universal bulk mass. 
Contours of EW constraints correspond to the  $65\%$ (dotted), $95\%$ (dashed) and $99\%$
(solid) C.L. fits for the oblique parameters.  
$\mu<0$ leads to KK-neutrino DM, which is unviable as a dark matter candidate. 
The $h_2$ resonance is included in `b', while only self-annihilation of KK photon
contributes in `a'.  Collider constraints shown are the current limits from ATLAS and CMS $W^\prime$ (dotted black), 
ATLAS dilepton (solid red), as well as the projected 10 fb$^{-1}$ reach (dotted red) of the ATLAS dilepton search.
Current CDF and ATLAS limit from dijet search are shown in dashed-magenta curves.}
\label{fig:muL}
\end{figure}

In Fig.~\ref{fig:muL}, we show the electroweak constraints in Split UED for a universal bulk mass by tracing the $65\%$ (dotted), $95\%$ (dashed) and $99\%$ (solid) C.L.~fit 
contours (in blue) in the $(\mu,\,R^{-1})$ space, 
which is consistent with results in Ref.~\cite{arXiv:1111.7250} 
\footnote{We represent our results in the ($\mu R$, $R^{-1}$) space for easy comparison with results in Ref. \cite{arXiv:1111.7250}.
However we find it more convenient to use $\mu L = \mu R \frac{\pi}{2}$ instead of $\mu R$. 
Therefore we use $\mu L$ in all figures except Fig. \ref{fig:muL}.}.  
We used the following experimental bounds on new physics contributions to
the oblique parameters: 
$S_{NP}=0.04\pm0.10$, $T_{NP}=0.05\pm0.11$, $U_{NP}=0.08\pm0.11$, 
for a reference point $m_{h}=120 \gev$ and $m_{t}=173 \gev$ with  correlation
coefficients  of $+0.89$ ($-0.45$, $-0.69$) between 
$S_{NP}$ and $T_{NP}$ ($S_{NP}$ and $U_{NP}$, $T_{NP}$ and $U_{NP}$), 
which have recently been updated by the \emph{Gfitter} collaboration \cite{ewGfitter}. 

Constraints on $R^{-1}$ become stronger than in MUED for a large value of $\mu$ ($\mu R \gsim 1$) 
due to enhancement in couplings of level-2$n$ gauge mode to SM fermion pair (see Eqs. \eqref{eq:g002n1}-\eqref{eq:g002n2}). 
It is interesting to notice that the lower bound on KK mass scale in MUED, $R^{-1}  \gsim 700$ GeV, is reduced for $\mu R < 1$.  
At $\mu R \sim 0.6$, the EW
constraints lead to the minimum allowed value of $R^{-1} \gsim 400$ GeV.

\subsection{Relic Abundance of KK Photon}
\label{sec:dm}

As discussed in the introduction, 
annihilation cross sections are reduced due to the heaviness of KK fermions in
the $t$- and $u$-channels, leading to larger relic density thus lower LKP mass,
to keep consistency with cosmological observations. 
Hence it is important in SUED to take into account different KK fermion masses, when computing relic abundance.
The masses of level-1 KK fermions are given by $m_1 = \sqrt{ k_1^2 + \mu^2 }$ in the absence of loop corrections ignoring EW symmetry breaking effect 
(see Eq. (\ref{eq:mf})). For $\mu \neq 0$, KK fermions are heavier than KK photon, $m_{f_1} > m_{\gamma_1} \approx R^{-1}$. 
Without an $s$-channel resonance, non-relativistic velocity expansion 
\begin{equation}
\sigma_{tree} v = a + b \, v^2 + {\cal O} \left ( v^4 \right )\, ,
\end{equation}
is a good approximation to obtain relic density \cite{servanttait} and 
we take two leading terms in the annihilation cross sections from Ref. \cite{Kong:2005hn}. 
For tree-level annihilation cross section via KK fermion exchange, they are
\begin{eqnarray}
\label{eq:aterm}
a &=& \sum_{f} \frac{32 \pi \alpha_Y^2 N_c m_{\gamma_1}^2}{9} 
  \left ( \frac{Y_{f_L}^4  }{(m_{\gamma_1}^2 + m_{f_{L1}}^2 )^2} 
         + \frac{Y_{f_R}^4  }{(m_{\gamma_1}^2 + m_{f_{R1}}^2 )^2} \right ) \, ,\\
b &=& - \sum_{f} \frac{4 \pi \alpha_Y^2 N_c m_{\gamma_1}^2 }{27} \Biggl(
   Y_{f_L}^4 \frac{11 m_{\gamma_1}^4 + 14 m^2 m_{f_{L1}}^2 - 13 m_{f_{L1}}^4 }{(m_{\gamma_1}^2 + m_{f_{R1}}^2 )^4} 
\nonumber 
\\
&&\qquad\qquad\qquad\qquad
 +Y_{f_R}^4 \frac{11 m_{\gamma_1}^4 + 14 m_{\gamma_1}^2 m_{f_{L1}}^2 - 13 m_{f_{L1}}^4 }{(m_{\gamma_1}^2 + m_{f_{R1}}^2 )^4}
\Biggr) \, ,
\label{eq:bterm}
\end{eqnarray}
where $\alpha_Y={g^\prime}^2/4\pi$ with $g^\prime$ being the $U(1)_Y$ hypercharge gauge coupling, and $N_c=3 \, (1)$ for $f$ being quark (lepton). 
$Y_f$ is the hypercharge of the fermion $f$. $f_{L1}$ ($f_{R1}$) represents $SU(2)_W$-doublet (singlet) KK fermion at level-1. 
The contribution from the Higgs final states remains the same as in MUED. 

Results from the relic abundance constraint ($\Omega h^2 = 0.1123 \pm 0.0035$) are shown in Fig.~\ref{fig:muL} in the SUED parameter space with a universal bulk mass, 
represented as the green band `a' (`b') without (with) resonance annihilation 
of KK photons through $h_2$~\cite{Kakizaki:2005en}.
The thickness in the green bands corresponds to uncertainty in $\Omega h^2$.  
In principle, $R^{-1}$ lower than the bands is still allowed for $\Omega h^2 < 0.1123$ but
other source of DM abundance is needed to account for the deficit. 
A 5D fermion mass parameter $\mu<0$ leads to a LKP which
is the KK partner of a SM neutrino and does not provide a viable
dark matter candidate \cite{sUED2,ST2}. 
Unlike the non-trivial behavior in oblique corrections, the lower bound on the preferred range of $R^{-1}$ from relic abundance decreases monotonically when $\mu$ increases.

There are two important corrections to the relic abundance of KK photon.  First,
coannihilation processes are important in MUED \cite{servanttait,Kong:2005hn} due
to the mass degeneracy.  Specifically, coannihilations with
$SU_W(2)$-singlet leptons reduce the LKP mass, while coannihilation
processes with other KK particles tend to increase it.  
Second, KK resonance at level-2 play an important role and increase the KK mass scale significantly \cite{Kakizaki:2005en}. 
In MUED, the mass splitting between level-2 KK Higgs and KK
photon is about 1-2\% for most of the allowed parameter space and the corresponding
enhancement in the relic abundance is $\sim$ 30\%.  The improved analysis
including coannihilation processes can be found in Ref. \cite{Kakizaki:2005uy}.
Resonance effects of other level-2 KK particles are studied in Ref.
\cite{Kakizaki:2006dz}.  
It is also  noticed in Ref. \cite{Belanger:2010yx} that
when allowing level-2 particles in the final state,
mainly $\gamma_2$ and $h_2$, 
the relic abundance decreases sharply, shifting the preferred value
of the dark matter mass above the TeV scale.
This is due to the important contribution of the coannihilation
channels ($\ell_1 \gamma_1 \to \ell \gamma_2$)
that are enhanced by the exchange near resonance of the level-2 KK singlet lepton. 
All these resonance effects tend to increase the mass scale of KK photon, while coannihilations with
$SU(2)_W$-singlet leptons tend to move in the opposite way.
Results in MUED from Ref. \cite{Belanger:2010yx}, for $m_h=120$ GeV, $\Lambda R=20$ and $R^{-1}=1$ TeV, show that 
the relative contributions to $\Omega h^2$ are
$\gamma_1 \ell_1 \sim 0.6$,
$\ell_1 \ell_1 \sim 0.13$,
$\gamma_1 h_1 \sim 0.09$,
$\gamma_1 \gamma_1 \sim 0.06$,
$\ell_1 h_1 \sim 0.05$,
$V_1 h_1 \sim 0.02$,
$h_1h_1 \sim 0.02$,
$\gamma_1 \ell_1 \sim 0.017$, and
$V_1 \ell_1 \sim 0.01$,
among which all fermion initial states are negligible for a sizable bulk mass in SUED.
The remaining important processes are then $\gamma_1 h_1$, $\gamma_1 \gamma_1$,
$V_1 h_1$ and $h_1 h_1$. Among them, $\gamma_1 h_1$ and $\gamma_1 \gamma_1$ 
are the dominant processes to determine the relic abundance.

In our study, we do not include coannihilation processes among KK fermions
and KK photon since there can be a relatively large mass gap  in the presence of a bulk mass. 
They become important only when $\mu R \lsim 0.01$, i.e., near the MUED limit.  
It is essentially the size of 1-loop radiative corrections in MUED, where
the correction to masses of ${\rm SU}(2)_W$-singlet KK leptons is $\sim$ 1\%. 
For the same reason, we do not include processes such as $\ell_1 \ell_1 \to Z_2 (\gamma_2)$, $\nu_1
\ell_1 \to W_2$, $\ell_1 \gamma_1 \to \ell_2 \ell_0$ etc.
However, we attempt to include some effects of $h_2$ resonance in $\gamma_1 \gamma_1 \to h_2$, although the bosonic sector may or may not stay the same as in MUED. 
Following the procedure described in Ref. \cite{Kakizaki:2005en}, we have numerically integrated the thermally averaged cross section including the Higgs resonance ($\sigma_{res}$) 
\footnote{We have used the decay width of the $h_2$ in MUED, which should be a good approximation for $\mu R <  1$. 
For the purpose of setting bounds on $\mu$, this is acceptable since LHC already constrain $\mu R < 0.2 \sim 0.3$, as we will see later.}. 
The improved relic abundance including $h_2$ resonance is labeled as `b' in Fig.  \ref{fig:muL}. 
Other coannihilation processes with KK bosons such as 
$\gamma_1 h_1^\pm \to ( W_2^\pm \, , h_2^\pm )$, $\gamma_1 h_1 \to  (A_2 \, , \gamma_2 \, , Z_2)$, 
$A_1 \,   A_1\, \to h_2$,   $h_1^+ \,  h_1^- \to ( h_2, Z_2, \gamma_2)$ 
etc may still contribute to the final relic abundance in principle \cite{Kakizaki:2006dz}. 
In the absence of a complete knowledge of the mass spectrum, 
our results are valid in the limit where all KK bosons other than the KK photon are
heavy and decoupled from the relic abundance calculation. 
These estimates should provide a ballpark range, since resonance effects in the
coannihilations are known to be less than about 30\% in MUED
\cite{Kakizaki:2006dz}.  One should revisit more systematically with radiatively
corrected KK masses, which is not known currently. 

The bulk mass parameter $\mu$ can also be constrained from below by dark matter direct detection experiments.
The main process involves the $s$- and $t$-channel exchange of KK quarks at level-1 ($q_1$) between KK photon $\gamma_1$ and the nucleus.
Current limit from XENON100~\cite{Aprile:2011hi} data implies a lower limit of $\mu R \gsim 0.01$, which is not shown in our plot.
At small $\mu_Q$, $q_1$ and $\gamma_1$ are nearly degenerate, and the direct detection cross section is
enhanced resonantly. Our calculation does not include a full treatment of radiative
corrections to the mass spectrum nor the finite width effect. 

In summary, coannihilation effect with KK leptons are negligible for a large bulk mass and
the effect with level-2 final states are expected to be small. 
We considered the Higgs resonance effect as shown in Fig. \ref{fig:muL}. 
This constraint on the dark matter abundance sets an upper limit on $R^{-1}$.

\subsection{Four Fermi Contact Interaction}
\label{sec:ff}

\begin{figure*}[t]
\centerline{    \epsfig{file=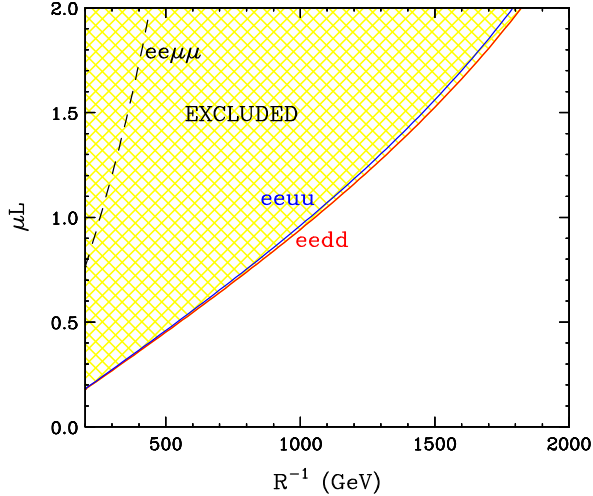, width=.5\textwidth}  }
\caption{\sl The four Fermi interaction excludes the upper-left corner of the $\mu L$-$R^{-1}$ plane.}
\label{fig:4fermi}
\end{figure*}

 In UED models, the Kaluza-Klein weak
gauge bosons can contribute to the four Fermi  contact operators, which can be
constrained by the experiments.  We consider lepton-lepton, lepton-quark
and quark-quark contact interactions which are described by the effective operators
 of the form of $\ell \bar{\ell}\ell\bar{\ell}$, $e\bar{e}q\bar{q}$ and
$q\bar{q}q\bar{q}$, respectively, with $\ell=e,\mu,\tau$  \cite{Cho:1997kf}.
It turns out the most stringent bound arises from the
electron-quark contact interactions of the form of  $e\bar{e}u \bar{u}$  and
$e\bar{e} d\bar{d}$ (see Table \ref{table:contact}) for $\mu_L=\mu_Q=\mu$ with $\eta_{AB}^q=\pm 1$: 
\begin{eqnarray}
{\cal L}^{eq}_{\rm eff} \ni  \sum_{q=u,d}\sum_{
\left\{A,B\right\}=\left\{L,R\right\}} \frac{4\pi } {\Lambda^2_{q, AB}}  \,  \eta^q_{AB} \, \bar{e}_A
\gamma^\mu e_A \bar{q}_B \gamma_\mu q_B \, , 
\end{eqnarray}
where 
\begin{eqnarray}
\frac{4\pi }{\Lambda^2_{q, AB}}  \eta^q_{AB} &=& 4\pi N_c \sum_{n=1}^{\infty}  \left({\cal F}_{0 0 }^{2n} (\mu  R)\right)^2 
                 \times \left[ \frac{3}{5}\frac{\alpha_1Y_{e_A} Y_{q_B} }{Q^2-M_{B_{2n}}^2} +\frac{\alpha_2T^3_{e_A} T^3_{q_B} }{Q^2-M_{W^3_{2n}}^2}\right]  \\
&\approx& -\pi N_c R^2 \left(\frac{3}{5}\alpha_1 Y_{e_A}Y_{q_B} +\alpha_2 T^3_{e_A} T^3_{q_B}\right)  
\times \sum_{n=1}^\infty \frac{ \left({\cal F}_{0 0 }^{2n} (\mu R)\right)^2 }{n^2}   \, ,
\label{eq:eeqq}
\end{eqnarray}
for a KK scale which is larger than the momentum transfer, $1/R\gg Q^2$.
Here $N_c=3$ is the color factor  ($N_c=1$ for $\ell \bar{\ell}\ell\bar{\ell}$ type contact interactions), and 
$Y$'s and $T$'s are the hypercharges and isospins of the corresponding fermions.  
We take $m_{B_{2n}}^2\approx m_{W_{2n}}^2 \approx (2n/R)^2$, considering $(m_W R)^2\ll 1$. 
The RG running effect of gauge couplings is included as well: 
\begin{eqnarray}
\alpha_1(\mu)=\frac{5}{3}\frac{g'^2(\mu)}{4\pi} = \frac{\alpha_1 (m_Z)}{1-\frac{b_1}{4\pi}\alpha_1(m_Z) \log \frac{\mu^2}{m_Z^2}},\\
\alpha_2(\mu)=\frac{g^2(\mu)}{4\pi} = \frac{\alpha_2 (m_Z)}{1-\frac{b_2}{4\pi}\alpha_2 (m_Z) \log \frac{\mu^2}{m_Z^2}},
\end{eqnarray}
with $\alpha_1(m_Z)\approx 0.017$,  $\alpha_2(m_Z)\approx 0.034$, and $(b_1 \, , b_2) = (41/10, -19/6)$.
We find that the contact interaction yields quite strong constraints for $1/R$ as shown in Fig. \ref{fig:4fermi}, 
and is comparable (slightly better large $R^{-1}$) to limits in the $W^\prime$ searches at the LHC. 
However we do not show this limit in Fig. \ref{fig:muL}, as the LHC limit is stronger. 
Note that $\mu L = \mu R \frac{\pi}{2}$.

\begin{table}[t]
\caption{Four Fermi contact interaction bounds in PDG(2010)  \cite{pdg}.}
\begin{center}
\begin{tabular}{c|c|c|c|c|c|c|c}
TeV&$eeee$& $ee\mu\mu$&  $ee\tau\tau$& $\ell\ell\ell\ell$&$qqqq$&$eeuu$&$eedd$ \\    
\hline
\hline
$\Lambda_{LL}^+$ &$>8.3$&$>8.5$&$>7.9$&$>9.1$&$>2.7$&$>23.3$&$>11.1$ \\
$\Lambda_{LL}^-$  &$>10.3$&$>9.5$&$>7.2$&$>10.3$&$2.4$&$>12.5$&$>26.4$
\end{tabular}
\end{center}
\label{table:contact}
\end{table}
%

\subsection{Anomalous Muon Magnetic Moment}
\label{sec:g-2}

In a previous study of MUED, 
the leading order correction to the anomalous magnetic moment of muon at 1-loop level is obtained \cite{Appelquist:2001jz}:
\begin{eqnarray}
\Delta a_\mu^{\rm MUED} \simeq \frac{\alpha}{8\pi} \sum_n \frac{(m_\mu  R)^2}{n^2} \left\{C_V + C_5 \right\} \, ,
\end{eqnarray}
where $m_\mu$ is the mass of muon and $C_\mu (C_5)$ stands for the effective
coupling of the vector (the fifth) components of the  KK gauge bosons and
Goldstone modes with the zero mode muon, which can be conveniently written separately:
\begin{eqnarray}
C_{V(5)}   &=& \left(C_{A^{\mu(5)}} + C_{Z^{\mu(5)}} + C_{W^{\mu(5)}}\right),\\
C_{A^\mu} &=& \frac{2}{3}=-\frac{2}{3}C_{A_5},\\
 C_{Z^\mu} &=&  -\frac{3+4 \sin^2 \theta_W \cos 2\theta_W}{3 \sin^2 2\theta_W},\\
 C_{Z^5}     &=& \frac{1+12\sin^2\theta_W \cos 2\theta_W}{6\sin^2 2\theta_W},\\
  C_{W^\mu} &=& 2 C_{W^5} = -C_{G^\pm}=-\frac{1}{3\sin^2\theta_W}.
\end{eqnarray}
We also include the Goldstone contribution ($C_{G^\pm}$). Taking $\alpha(m_Z)=1/127$ and $\sin^2 \theta_W=0.2316$, we get
$\Delta a_\mu^{\rm MUED}\approx -1.2 \times 10^{-11}$ for $R^{-1}=1$ TeV, which
is far below the detectable range.  Given that KK fermions are heavier in 
SUED, the contributions are  doubly suppressed by the fermion masses and by the
coupling constants ($\sim g_{\rm SM}{\cal F}_{n0}^n$). We found that
the contribution from the KK number conserving interactions in 
SUED could be reduced down to 61\% (17\%) of the value in MUED when $\mu_L L =$1.0 (3.0).

However, in SUED there are potentially more important KK number
violating contributions from loops containing the $(2n)^{\rm th}$ neutral KK gauge bosons, 
i.e., $B_{2n}$ and $W^3_{2n}$, as the zero mode muon has sizable
couplings to them ($\sim g_{\rm SM} {\cal F}_{00}^{2n}$):
\begin{eqnarray}
\Delta a_{\mu}^{B_{KK}}      &&\simeq -\frac{3\alpha_1}{16\pi} \cdot (m_\mu R)^2 \sum_n \left(\frac{{\cal F}_{00}^{2n}}{n}\right)^2 \times  L_1\, , \\
\Delta a_{\mu}^{W^3_{KK}} && \simeq -\frac{\alpha_2}{16\pi} \cdot (m_\mu R)^2 \sum_n \left(\frac{{\cal F}_{00}^{2n}}{n}\right)^2 \times  L_2 \, .
\end{eqnarray}  
Here $L_i=\frac{1}{2}\int_0^1 dx \frac{Q_i(x)}{1-x+ \lambda_n^2 x^2}$ is the loop functions with 
$\lambda_n =\frac{m_\mu}{m_{2n}}\approx \frac{m_\mu R}{2n}$, 
$Q_{B_{2n}}(x)=Q_V(x)$  and $Q_{W^3_{2n}}(x)=Q_V(x)-Q_A(x)$.  The vector and axial vector  coupling functions are
$Q_V(x)=2x^2(1-x)$ and $Q_A(x)=2x(1-x)(x-4)-4\lambda_n^2 x^3$, respectively. 
Explicitly, $L_1=1/3$ and $L_2=2$ when we take $\lambda_n=0$ for the integration
\cite{Jegerlehner:2009ry}. Finally, the leading contribution 
\footnote{We thank Tom Flacke for pointing out that 
there are other contributions which arise due to $f_2$-$f_2$-$V_{2n}$, $f_0$-$f_1$-$V_{2n+1}$, $f_0$-$f_3$-$V_{2n+1}$, etc. 
Individual contributions are expected to be smaller than the leading contribution due to the heaviness of KK fermions. 
However, summing over all possible KK fermion states, the net contribution may appear divergent.  
Often the sum is truncated by including KK states up to the corresponding cut off scale times the radius. 
With this, we think that the total contribution will not change by more than an order of magnitude and will be still below current experimental sensitivity.
This issue needs further investigation. }
from the neutral KK bosons is estimated as 
\begin{eqnarray}
\Delta a_\mu^{SUED} \approx -1.8 \times 10^{-11} \left(\frac{1~{\rm TeV}}{R^{-1}}\right)^2  \sum_n \left(\frac{{\cal F}_{00}^{2n}}{n}\right)^2 , ~~
\end{eqnarray}
where $\sum_n \left(\frac{{\cal F}_{00}^{2n}}{n}\right)^2<2$ for $\mu_L L<5$, 
which is still too small to be detected. This new contribution vanishes in 
the MUED limit where ${\cal F}_{00}^{2n}\to 0$. 

\subsection{Collider Bounds}
\label{sec:collider}
In MUED, the first level of KK excitations can be produced in pairs at
colliders, with the typical missing energy signature due to conservation of KK parity. 
Recent studies indicate a bound of $1/R\gsim 700 \gev$
\cite{collbounds} for MUED from the first year LHC data.  
While KK parity in UED is conserved, KK number is broken through 
loop-generated couplings between level-2 KK bosons and SM fermion pairs.
These KK bosons can appear as resonances in the dilepton or dijet final states, 
but their productions are heavily suppressed by loop factors.

In SUED, however, these couplings exist at tree-level in the presence of $\mu$.
Therefore their LHC limits are expected to be stronger than those in MUED.
Parameters $\mu$ and $R^{-1}$ can be constrained by searches in the
dijet~\cite{ATLAS2j,CMS2j}, dilepton~\cite{ATLAS2l,CMS2l} and 
$W^\prime$ (lepton + neutrino)~\cite{ATLASwp,CMSwp} channels.
Previous studies~\cite{Chen:2009gz,sUED2, SUED3} have explored this aspect in
some depth. In particular, the dilepton reach and exclusion are
mapped~\cite{sUED2} in the ($\mu$, $R^{-1}$) space for a 10 TeV LHC.
 
\begin{figure*}[t]
\centerline{
\epsfig{file=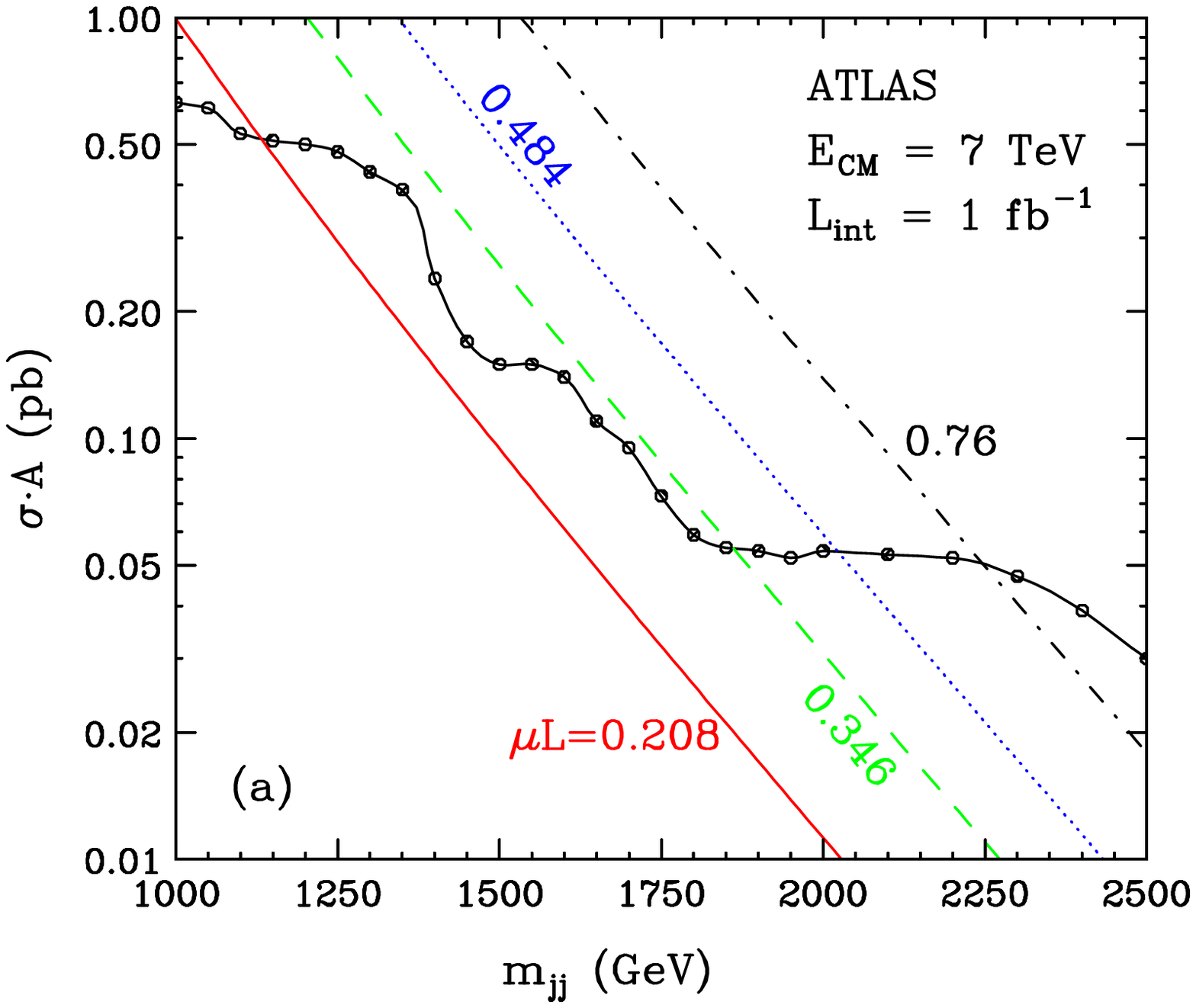, width=.33\textwidth} 
\epsfig{file=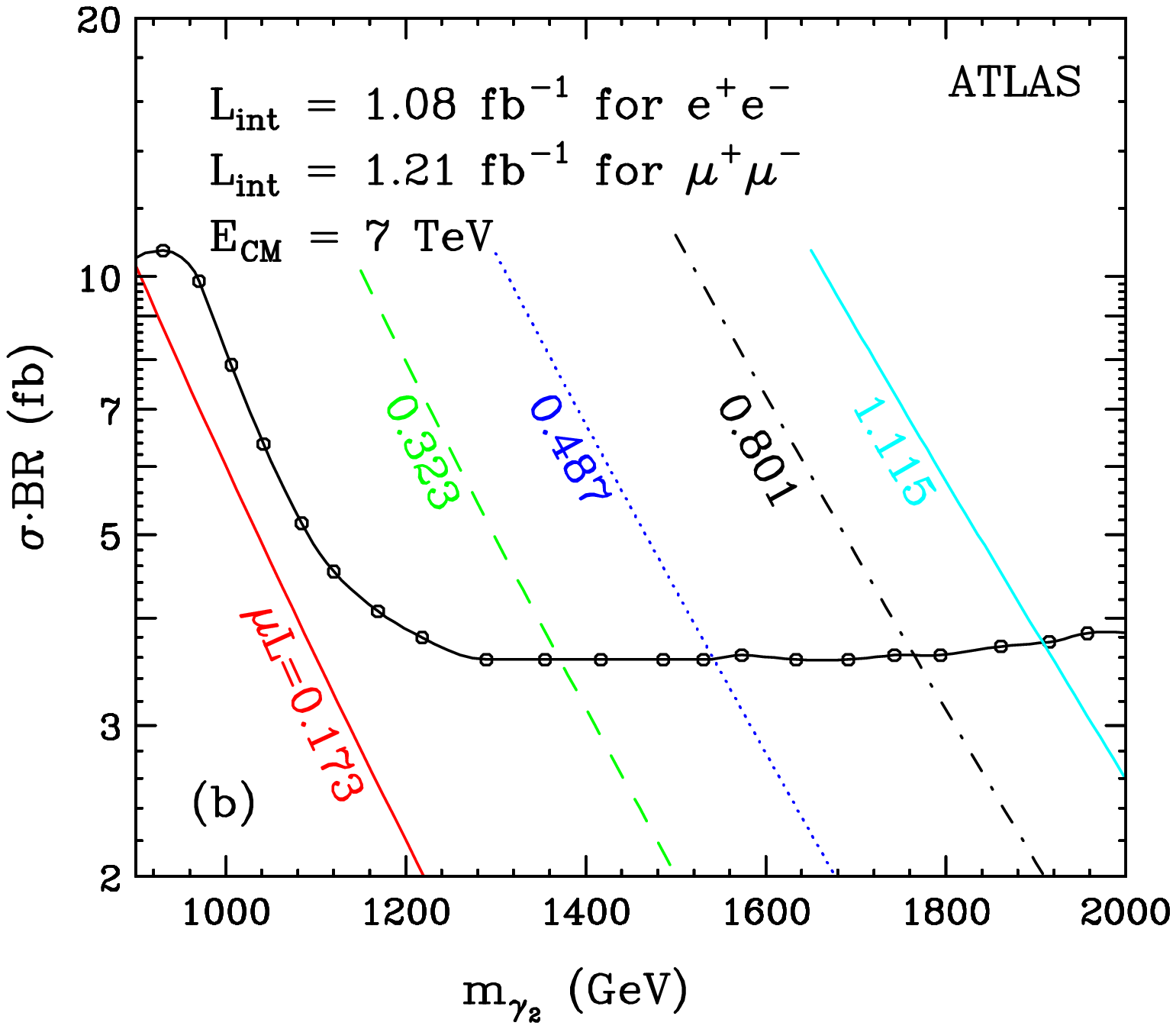, width=.32\textwidth} 
\epsfig{file=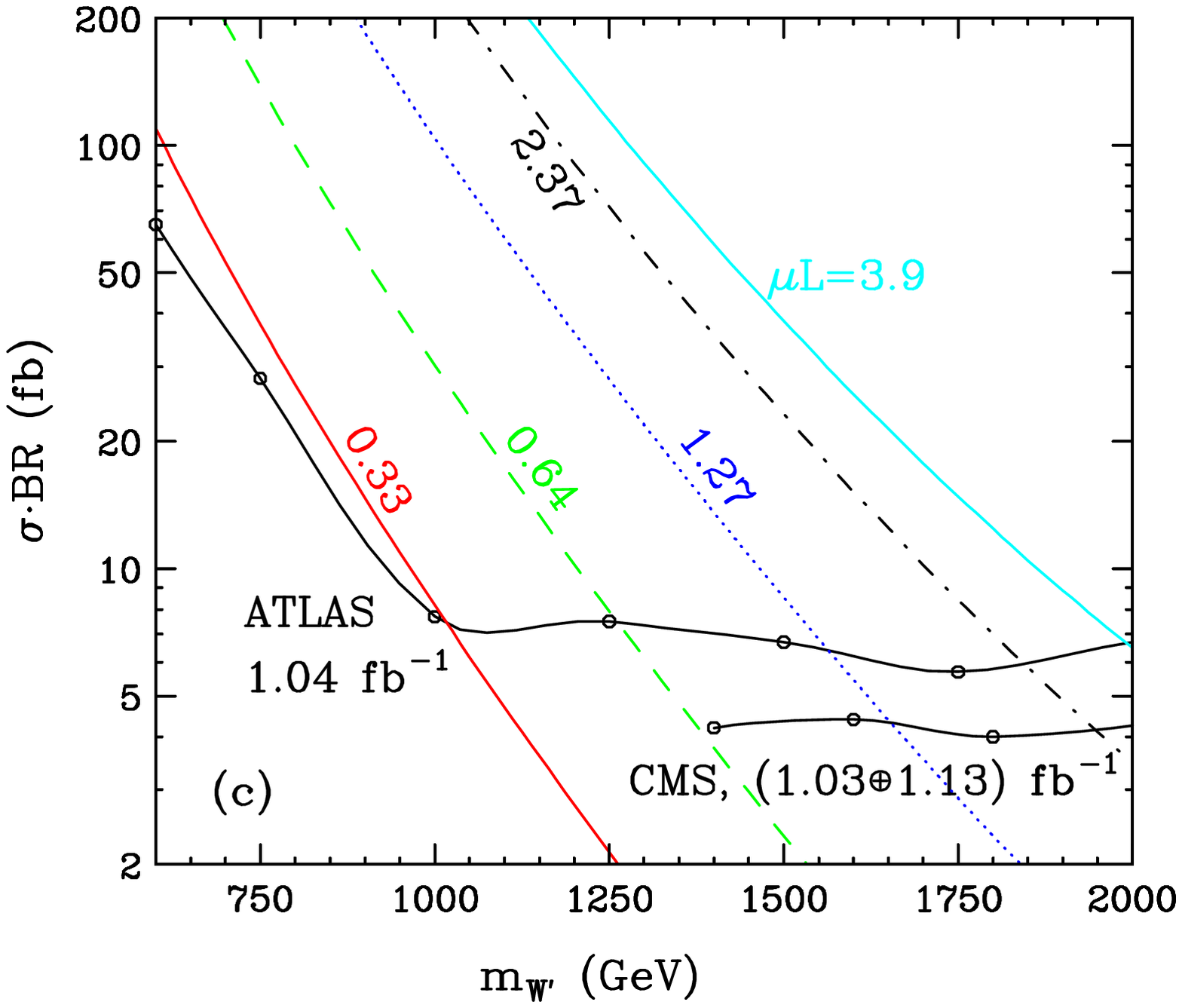, width=.33\textwidth}  }
\caption{\sl Bounds on masses of KK resonances in dijet ($jj$ in (a)), dilepton ($\ell\ell$ in (b)), and lepton plus missing momentum ($\ell\nu$ in (c)) channels.
Curves with circles represent the 95\% C.L. upper limit on signal cross sections.} 
\label{fig:collider}
\end{figure*}

We use CalcHEP~\cite{hep-ph/0412191} and CTEQ 5M PDF to evaluate cross sections. 
Appropriate cuts and efficiencies have been applied, following experimental studies \cite{ATLAS2j,CMS2j,ATLAS2l,CMS2l,ATLASwp,CMSwp}.
More accurate calculation depends on detailed mass spectrum of the model, which
is currently not known for SUED.  
We here adopt the same mass splittings as in MUED, with 
$M_{G_2} : M_{W_2} : M_{Z_2} : M_{\gamma_2} \approx 1.3 : 1.07 :1.07 : 1$, and 
$M_{\gamma_2} \approx \frac{2}{R}$. 
Widths of these resonances are computed automatically in CalcHEP.
Since all KK fermion final states are
prohibited due to their heaviness in most of the parameter space, 
we only consider SM fermion pairs in the final states. 
Our results on collider bounds are shown in Fig. \ref{fig:collider}, with  dijet in (a), dilepton in (b), and  lepton plus missing momentum in (c).
Curves with dots (black-solid) represent the 95\% C.L. upper limit on signal cross sections as  functions of the relevant resonance mass, 
while all other curves are signal cross sections in specific channels for various values of bulk mass.

The dijet channel includes resonances through  $G_2$, $\gamma_2$, $Z_2$ and
$W_2$, with a mass spread of $\sim$30\% of $m_{\gamma_2}$.
Production and hadronic decays of different resonances scale the same way with
the quark bulk mass $\mu_Q$ (but not $\mu_L$). 
For simplicity, we only consider $G_2$, which dominates the others with
strong  QCD couplings, as well as a larger mass where the background is smaller. 
We find that in SUED the ATLAS reach is slightly more sensitive to the dijet resonance than the CMS one, 
therefore we only include the ATLAS analysis below. 
In Fig. \ref{fig:collider}(a) we show the 95\% C.L. upper limit on cross section times acceptance ($\sigma \times {\cal A}$) 
as a function of the dijet resonance mass. 
We follow the procedure described in Ref. \cite{ATLAS2j} to set limits on the mass of the dijet resonance.
Acceptance is obtained by imposing the suitable kinematic cuts on $\eta$, $p_T$, $|\Delta \eta|$, and invariant mass ($m_{jj}$) employed in the ATLAS analysis with 1 fb$^{-1}$. 
We also include a factor of 0.92 to account for an approximate reduction of acceptance due to the calorimeter readout problem 
in the region of $\eta \in (-0.1, 1.5)$ and $\phi\in(-0.9, -0.5)$. 
To compare the 95\% C.L. upper limit and our signal cross section, 
the ratio of the mean mass and the standard deviation of the Gaussian resonance
is chosen as 5\%, which is the closest value for SUED.
Finally the bounds on the dijet invariant mass, $m_{jj} = 1.3 \frac{2}{R}$, can be read off from intersections in Fig. \ref{fig:collider}(a), and results are translated 
into the $(\mu R, R^{-1})$ plane, as shown as the magenta-dashed curve in Fig. \ref{fig:muL}. Note that $\mu L = \mu R \frac{\pi}{2}$. 

For the dilepton channel, we include both $\gamma_2$ and $Z_2$ resonances in estimating signal cross sections. 
$\mu_Q$ is important for the production while $\mu_L$ is relevant for the decay, although we are considering the universal case in this section. 
Fig. \ref{fig:collider}(b) shows the upper limit as a function of dilepton resonance mass ($m_{\ell\ell}$) as well as signal cross sections for various choices of $\mu$. 
Note that there are two resonances, $\gamma_2$ and $Z_2$, with a mass splitting of $m_{Z_2} = 1.07 m_{\gamma_2} = 1.07 \frac{2}{R}$, and the $x$-axis represent $m_{\gamma_2}$ in Fig. \ref{fig:collider}(b). 
We use results from ATLAS  with an integrated luminosity of 1.08 fb$^{-1}$ in the dielectron channel and  
1.21 fb$^{-1}$ in the dimuon channel \cite{ATLAS2l}. Above a resonance mass of 1.2 TeV, their limits stay constant.
Translating results to the $(\mu R, R^{-1})$ plane, we find that current LHC bounds in the dilepton channel (red-solid) is similar 
to that in the dijet channel (magenta-dashed), as shown in Fig. \ref{fig:muL}. 
However, as discussed in Ref. \cite{2ndKKrefs}, the bounds on SUED in the dilepton channel can be improved significantly by including indirect processes, 
which require a complete knowledge of mass spectrum. 
The projected bounds assuming 10 times more LHC data are shown as the red-dotted curve.

Finally bounds from the $W^\prime$ search in the lepton plus missing energy channel are shown in Fig. \ref{fig:collider} for both ATLAS and CMS, 
and their corresponding constraints on $( \mu$ and $R^{-1} )$ are shown in Fig. \ref{fig:muL} (black-dotted). 
CMS limits are slightly better while ATLAS covers lower mass region.
All collider bounds (dijet, dilepton and lepton plus missing energy) constrain
regions with large bulk mass and small KK mass scale.
Hence the upper-left corner of each curve is ruled out by these searches.

One of our main results is that considering oblique corrections, collider bounds and relic abundance constraints together, 
SUED parameter space is restricted to a region of  $650 \gev \lsim 1/R \lsim 850 \gev$ and $\mu R \lsim 0.2$ without the resonant annihilation (green band in the left).
With the resonance, we are restricted to $750\lsim 1/R \lsim 950 \gev$ and $\mu R \lsim 0.3$ (green band in the right).
In general, $\Omega h^2 < 0.1123$ is still acceptable, in which case the yellow-shaded region is allowed. 
Considering current performance of the LHC, the remaining parameter space of the universal bulk-mass will be highly constrained. 
For example, the dilepton resonance search with 10 fb$^{-1}$ fb will be sensitive down to $\mu R \lesssim $ 0.1-0.2.

\section{A Non-Universal Case: $\mu_L\neq \mu_Q$}
\label{sec:nonuniversal}
Once we remove the universal bulk mass requirement, 
fermions can take on more generic flavor and chiral structures.
To avoid stringent bounds from the flavor changing neutral current (FCNC) effects,  and for simplicity, 
we consider the case where all leptons have the same bulk mass $\mu_L$ and all quarks $\mu_Q$. 
The constraints on the
full parameter space of such model, ($\mu_L,\, \mu_Q,\, R^{-1}$), 
are considered in this section.

With different mass parameters $\mu_L$ for leptons and $\mu_Q$ for
quarks, the bound on $\mu_Q$ can be substantially weakened, because for
$\mu_L=0$, the couplings of the muon to non-zero KK $W$ modes vanish, and muon
decay only proceeds via the $W$ zero-mode. 
 In fact, for  non-universal
masses $\mu_L\neq\mu_Q$,  leptonic and hadronic channels at LEP are not
affected universally any more,  so that a treatment in terms of the oblique
$(S,T,U)$ parameters is insufficient, and a global fit to the LEP data is
required for a reliable electroweak analysis, which we reserve for future study.  However  we  still can find quite tight bounds on the parameter space 
by considering the potentially relevant constraints coming from corrections to
the decay rate of muon in  terms of Fermi constant ($\delta G_F$), and from the four
Fermi contact interactions of various types.  We also consider the anomalous
magnetic moment of muon, which currently is known to have a slight disagreement between the theoretical and experimental values.

\subsection{Electroweak Precision Measurements: Oblique Corrections, Four Fermi Operators and Anomalous Magnetic Moment}
\label{sec:stu2}

Unlike the universal case ($\mu_L = \mu_Q$)  in section \ref{sec:stu}, 
the analysis with oblique parameters $(S,T,U)$
may not provide enough information on the new physics effects as
non-universal corrections to couplings are involved.
However, we still find a stringent bound in the non-universal case using an effective parametrization of
oblique corrections as follows: For $(S,T,U)_{\rm UED}$, the loop
contributions from KK tops are again the most relevant among all fermion contributions due to the fermion mass
suppression in Eq. \eqref{eq:replace}. The most important constraint
for $\delta G_F$ is from the precise measurement of the muon decay process, for
which  the leading order correction vanishes when $\mu_L=0$ regardless the
value of $\mu_Q$. Capturing these facts we conveniently replace $\mu \to \mu_Q$
for $(S,T,U)_{\rm UED}$ and $\mu \to \mu_L$ for $\delta G_F$ as
leading order contributions. 
Then we use the oblique parameter fits and correlation matrix from the \emph{Gfitter} collaboration \cite{ewGfitter}. 
\begin{figure*}[t]
\centerline{
\epsfig{file=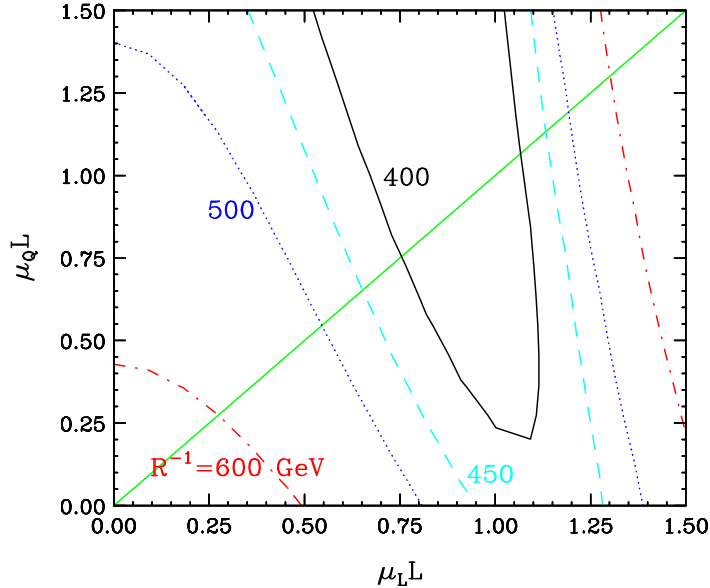, width=.6\textwidth} }
\caption{\sl Oblique corrections in the case of two fermion-bulk masses. 
Each contour disfavors its outside region for a given $R^{-1}$ (99\% C.L.). The universal case is along the green line ($\mu_L=\mu_Q$).} 
\label{fig:STU_2}
\end{figure*}

Fig.~\ref{fig:STU_2} shows contours of $R^{-1}$ that are consistent with 
EW observables in the $(\mu_L,\mu_Q)$ plane at 99\% C.L.
Each contour disfavors its region outside.  
The universal bulk-mass case can be obtained by taking limits along the diagonal line (in green). 
For instance, taking the contour of $R^{-1}=500$ GeV, one can read off the allowed ranges of $\mu_Q L=\mu_L L$  
$\in$ [0.545, 1.192] ($\mu_Q R=\mu_L R  \in$ [0.347, 0.759]),
which is consistent with the universal case as shown in Fig. \ref{fig:muL}. 

\begin{figure*}[t]
\centerline{
\epsfig{file=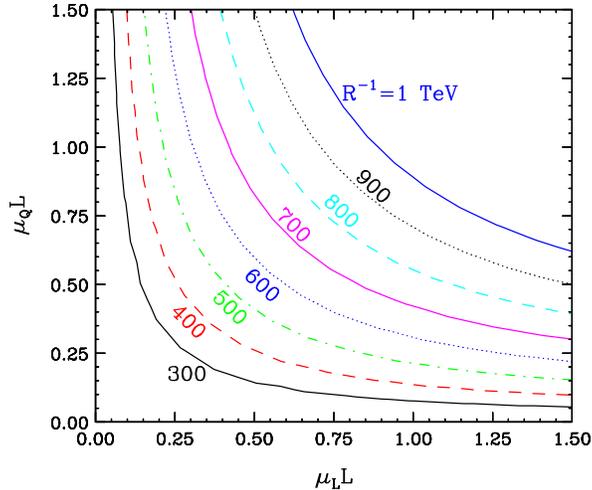, width=.5\textwidth} }   
\caption{\sl Four Fermi Interaction (from $eedd$) constrains the upper-right corner for a given $R^{-1}$. }
\label{fig:4F}
\end{figure*}

As oblique corrections may not capture the full features of corrections, we also wish to include four Fermi contact operators. 
Actually the tree level $0$-$0$-$(2n)$ interactions could be large enough in the presence of bulk masses to greatly enhance the contact interactions. 
To analyze four Fermi operators, we can generalize Eq. \eqref{eq:eeqq} for $\mu_L \neq \mu_Q$ case, simply by replacing $\left({\cal F}_{0 0}^{ 2n} (\mu R)\right)^2$ with ${\cal F}_{0 0}^{ 2n} (\mu_Q R) {\cal F}_{0 0}^{ 2n} (\mu_L R)$.
${\cal F}^{2n}_{0 0}$ increases as $\mu$ increases so that the upper-right corner of
$(\mu_L, \mu_Q)$ plane is constrained by the four Fermi interactions. 
In Fig. \ref{fig:4F} we show bounds from $eedd$ operator, which is the most stringent among all operators we consider (see Table \ref{table:contact}). 
For a given point $(\mu_L, \mu_Q)$, one can read off lower bound on $R^{-1}$.
For example, for $\mu_L L\approx 1 \approx \mu_Q L$, $R^{-1}\lesssim 1$ TeV is ruled out but if one of the
bulk mass parameter becomes small a quite large portion of the parameter space is still available.

In the leading order approximation, the anomalous muon magnetic moment is only sensitive to $\mu_L$ so that the non-universal case with $\mu_L \neq \mu_Q$ does not add more information other than that  in section~\ref{sec:g-2}.

\subsection{Relic Abundance of KK Photon}
\label{sec:dm2}

\begin{figure*}[t]
\centerline{
\epsfig{file=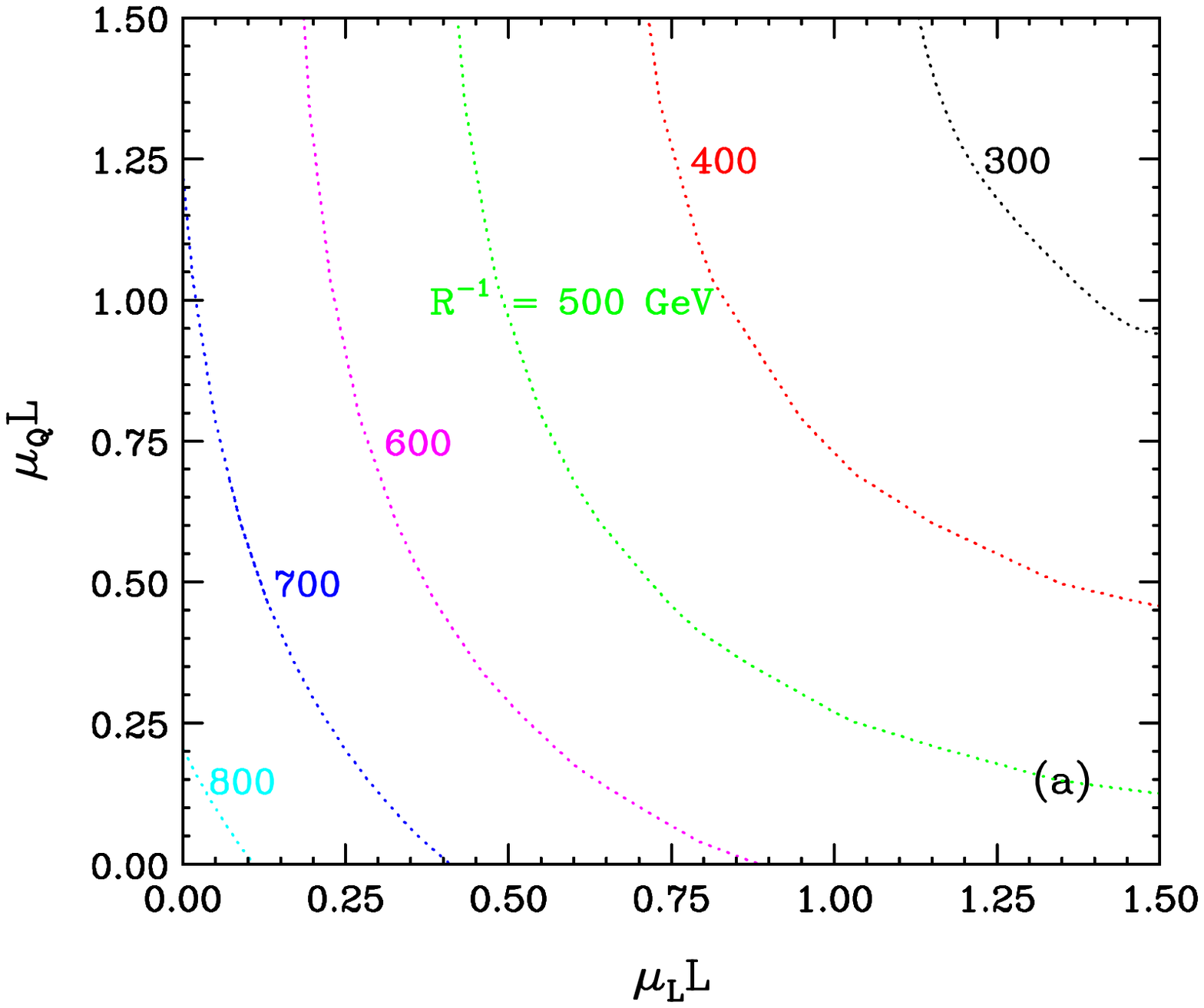, width=.5\textwidth} 
\epsfig{file=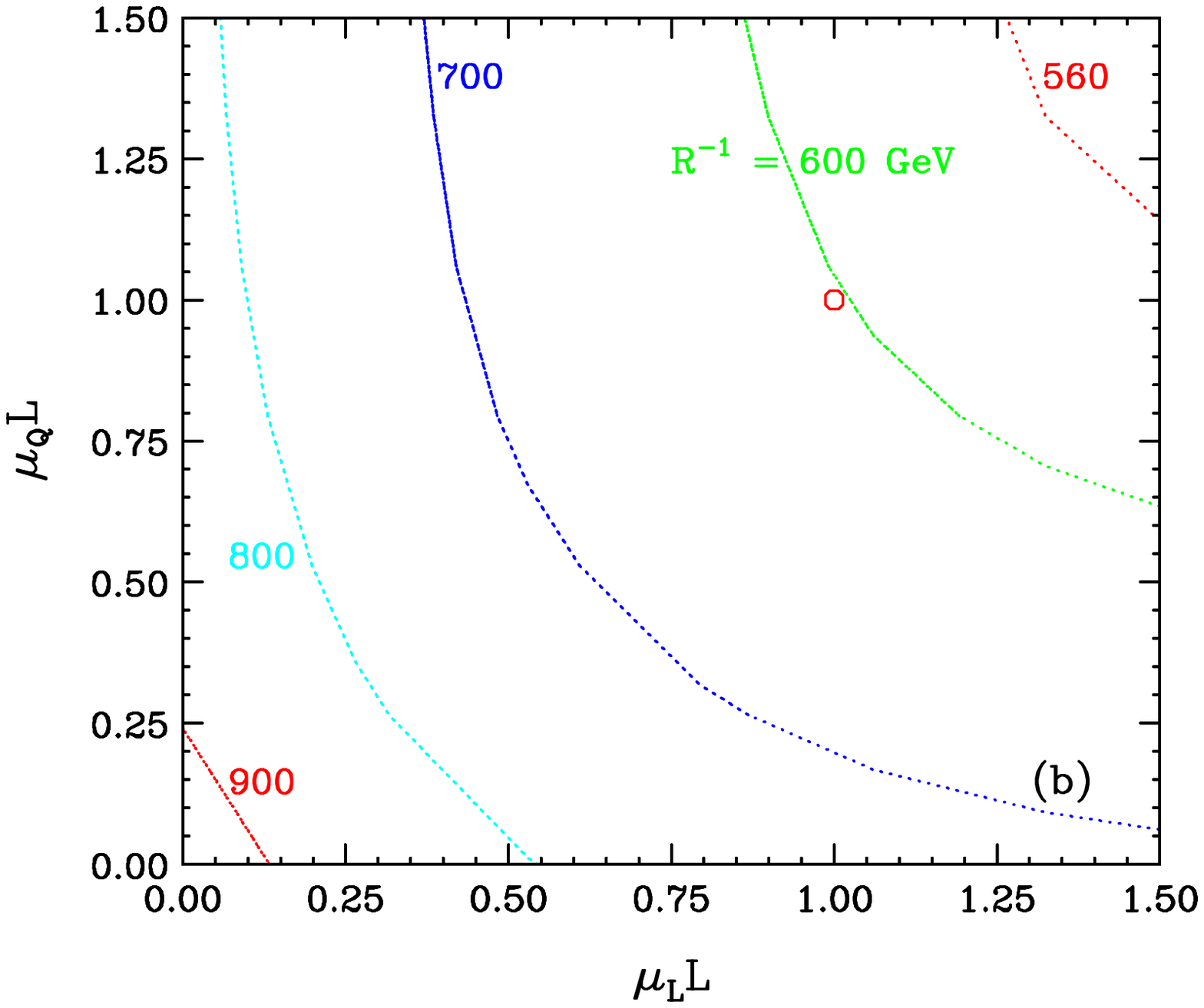, width=.5\textwidth}   }
\caption{\sl 
Relic abundance for two bulk masses, $\mu_Q$ and $\mu_L$ for quark and lepton sector, respectively. 
Contours represent values of $R^{-1}$ that lead to $\Omega h^2 = 0.1123$ 
 without (with) $h_2$ resonance in (a) (in (b)). 
For a given set of ($\mu_L$, $\mu_Q$), the contours serve as an upper bound on allowed $R^{-1}$.} 
\label{fig:oh2}
\end{figure*}

We show our results on relic density in Fig. \ref{fig:oh2}, where contours
represent values of $R^{-1}$ that lead to $\Omega h^2 = 0.1123$ without (with)
$h_2$ resonance in (a) (in (b)).  
In the calculation of relic abundance, two bulk masses factor in differently in each fermion sector.
Requiring that KK photon accounts for all of the dark matter in our universe 
($\Omega h^2 = 0.1123$) leads to indirect exclusion. 
KK photon is mostly the KK partner of the hypercharge gauge boson and couples to leptons stronger than to quarks due to larger hypercharges of leptons. 
We therefore notice that $\mu_L$ is more constrained than $\mu_Q$.  

For a given set of ($\mu_L$, $\mu_Q$), the contours serve as an upper bound on allowed
$R^{-1}$.  Any value larger than $R^{-1}$ would have a problem with overclosure.
For instance, $( \mu_L L , \mu_Q L) = (1,1)$ denoted by a circle (in red) in Fig.
\ref{fig:oh2}(b) is allowed and the corresponding value of contour is about
600 GeV for $\Omega h^2 = 0.1123$.  Therefore we learn that the model point,
$( \mu_L L , \mu_Q L) = (1,1)$, is not allowed for $R^{-1} > 600$ GeV.  
In principle, $R^{-1} < 600$ GeV for $( \mu_L L , \mu_Q L) = (1,1)$ can still be 
consistent with cosmological observation but 
one needs multiple dark matter candidates to make up the difference.  
\begin{figure*}[t]
\centerline{
\epsfig{file=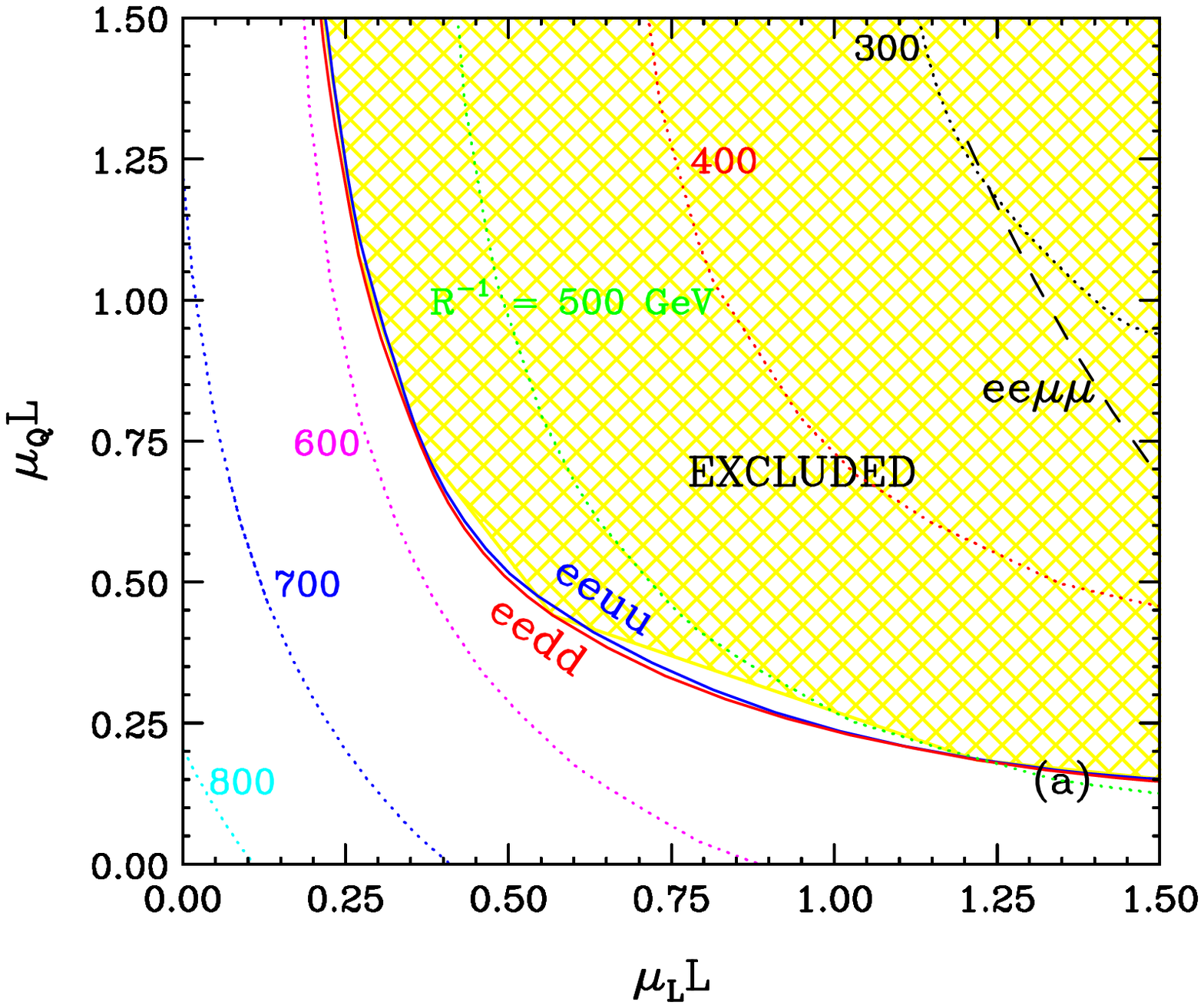, width=.5\textwidth} 
\epsfig{file=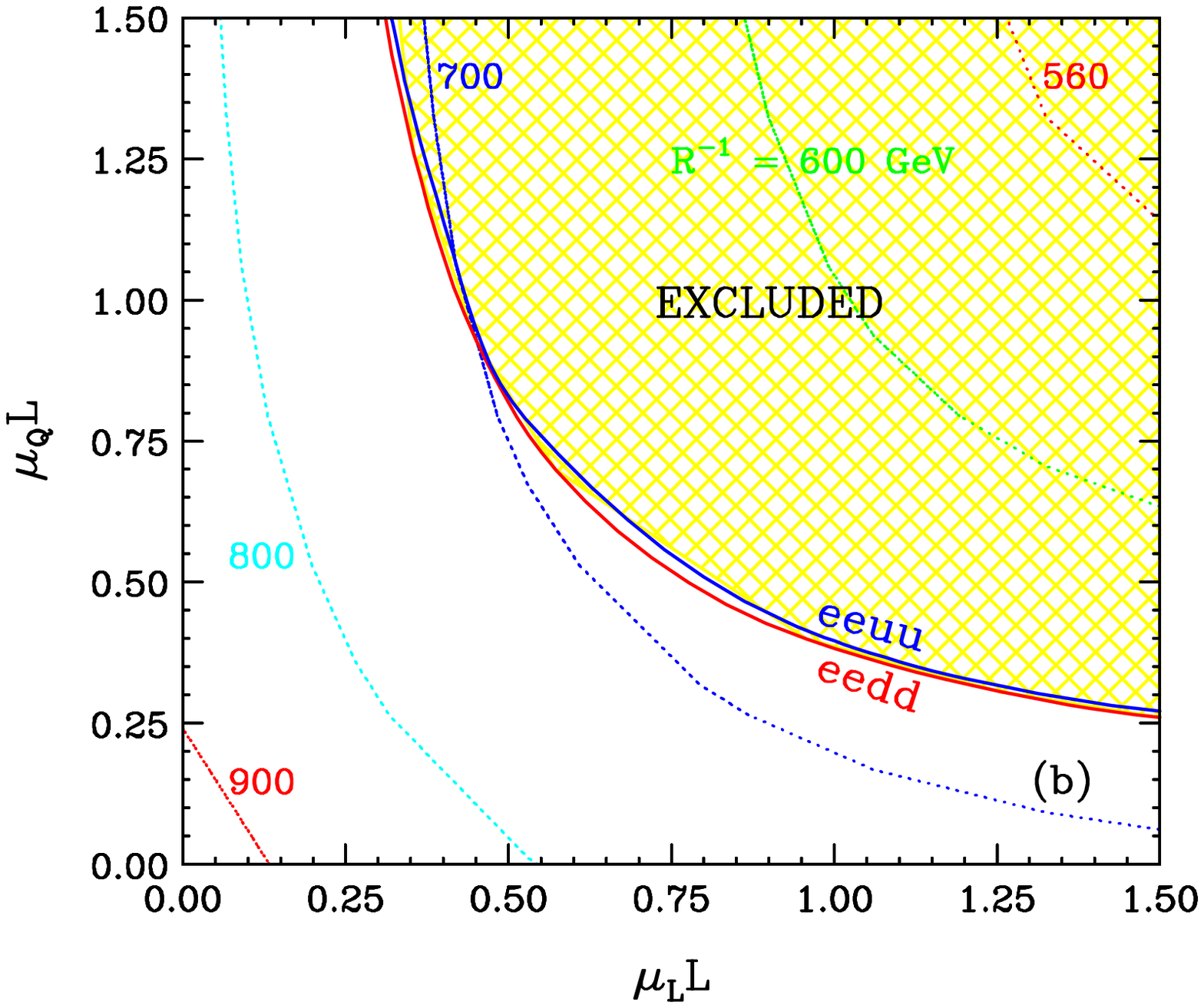, width=.5\textwidth} }
\caption{\sl Four Fermi Interaction combined with relic density without (with) $h_2$ resonance in (a) (in (b)).} 
\label{fig:4F2}
\end{figure*}
In Fig. \ref{fig:4F2} we depict the bounds from the four Fermi contact interactions 
(especially the operators $eedd, eeuu, ee\mu\mu$) combined with relic density without (with) $h_2$ resonance.
A slightly larger parameter space survives when the $h_2$ resonance is effective.

\subsection{Collider Bounds}
\label{sec:collider2}
In the non-universal bulk mass case, 
both dilepton and $W^\prime$ channels constrain the model
in the full $(\mu_Q,\,\mu_L,\,R^{-1})$ parameter space, 
while the dijet channel only constrain $(\mu_Q,\,R^{-1})$. 
Since the $W^\prime$ bound is weaker than the dilepton one as demonstrated in the universal bulk mass case in section \ref{sec:collider},
we here consider the dijet and dilepton channels only. 

We would like to emphasize that when the bulk lepton masses are vanishingly small $\mu_L R\ll 1$,  
the second KK modes of the electroweak gauge bosons cannot contribute to
$s$-channel exchange at tree-level in the dilepton channel, which serves as one of the main
search channels for UED and its extensions at the LHC.

\begin{figure*}[t]
\centerline{
\epsfig{file=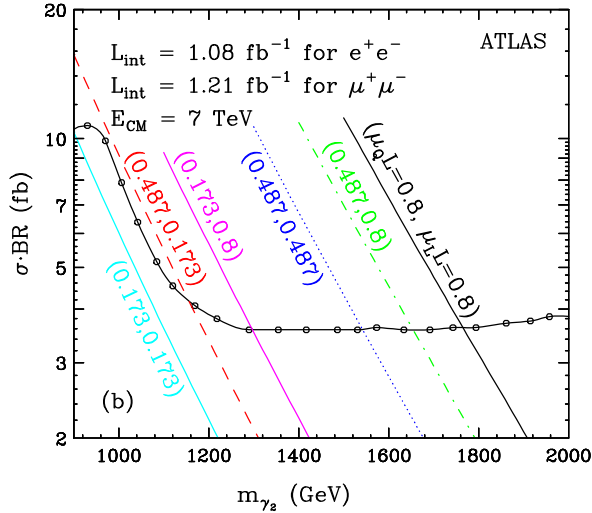, width=.5\textwidth}  }
\caption{\sl Bounds on the masses of KK resonances in the dilepton ($\ell\ell$) channel.
The curve with dots represents the 95\% C.L. upper limit on signal cross sections.} 
\label{fig:collider2}
\end{figure*}

\begin{figure}[t]
\centerline{
\epsfig{file=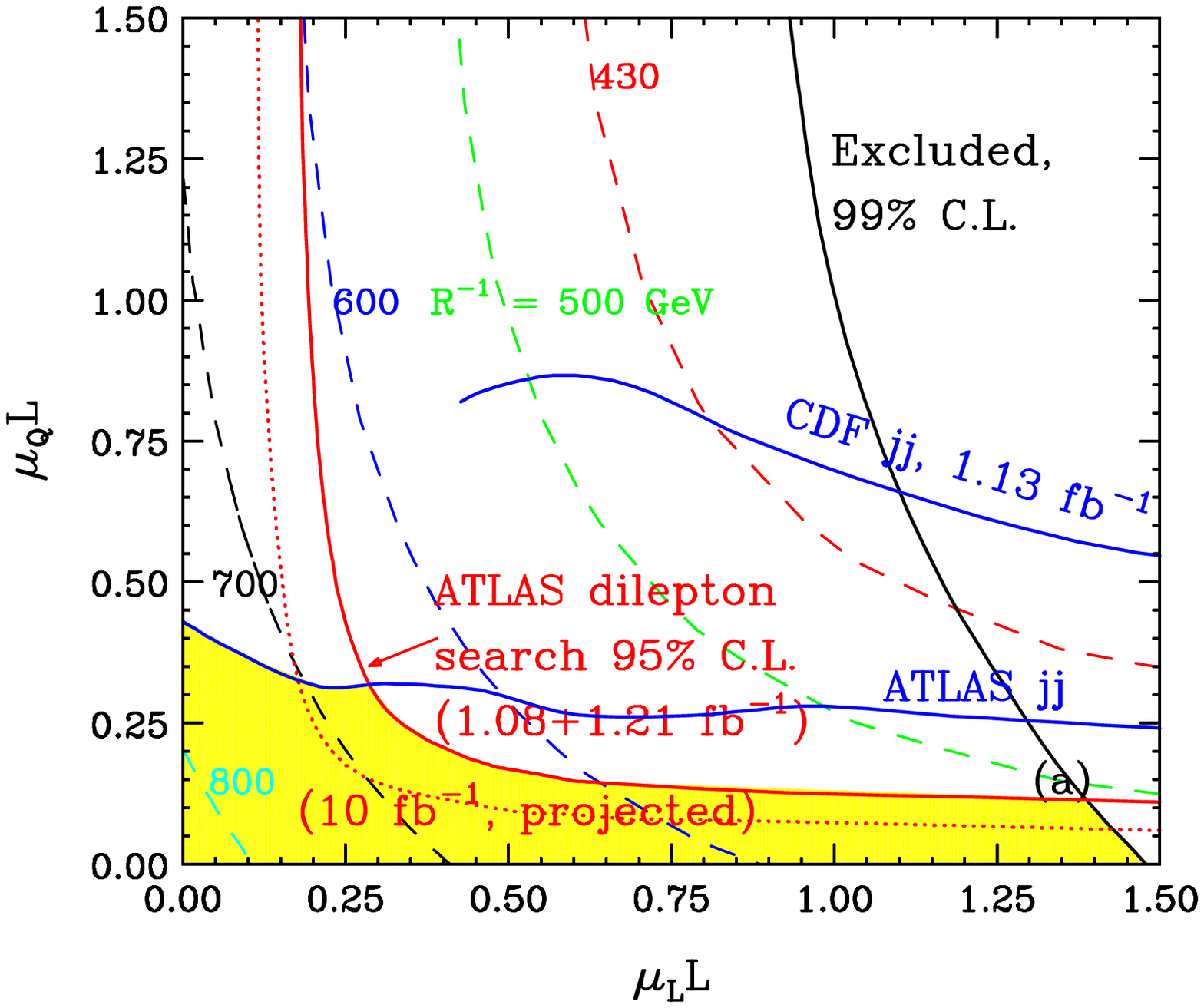,         width=.5\textwidth}
\epsfig{file=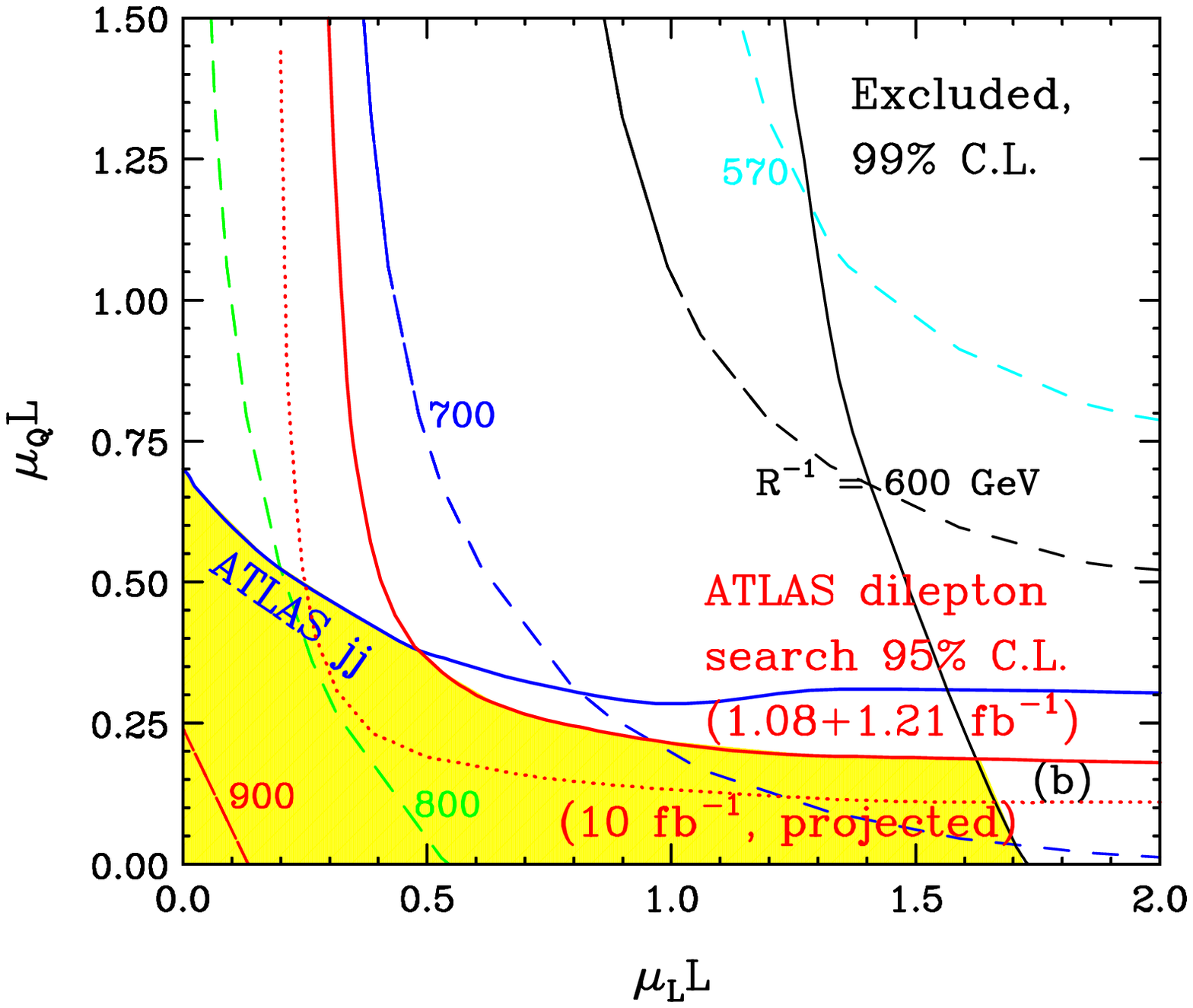, width=.5\textwidth}  }
\caption{\sl Relic abundance, electroweak and collider constraints for two bulk masses. 
Contours represent values of $R^{-1}$ that lead to $\Omega h^2 = 0.1123$ 
without (with) $h_2$ resonance in (a) (in (b)).
Region in the right side of black curve (labeled as ``Excluded") is disfavored by oblique corrections (99\% C.L.) for
$\Omega h^2 = 0.1123$. 
The red curves represent exclusion limit from current (solid) and projected 
10 fb$^{-1}$ (dotted) ATLAS dilepton search. The dijet constraints are shown in blue-solid curves. 
Dashed curves represent contours of $R^{-1}$ that is consistent with $\Omega h^2 = 0.1123$. Note that $L=\pi R/2$.}
\label{fig:muQmuL}
\end{figure}

In Fig. \ref{fig:collider2}, we show the 95\% C.L. upper limit on signal cross sections (black, solid with dots),
as well as signal cross sections for various choices of ($\mu_Q, \mu_L$).
As expected, dilepton bounds in this case can be stronger or weaker depending on the magnitude of $\mu_Q$ and $\mu_L$.
For instance, for a fixed value of $\mu_Q L=0.487$, 
the corresponding mass bound for 
increasing $\mu_L L$ (from 0.173 to 0.487, 0.8) also grows stronger (from 1150 to 1550, 1650 GeV) 
(see three curves in red-dashed, blue-dotted, and green-dot-dashed for comparison). 
The dilepton limit, on the other hand,  is in fact a 2-dimensional `exclusion surface' in
the ($\mu_Q,\,\mu_L,\,R^{-1}$) space.
We recall that the relic abundance requirement $\Omega h^2 = 0.1123$
also corresponds to a 2-dimensional `constraint surface'.
By intersecting these two surfaces and projecting down to the $(\mu_Q,\,\mu_L)$
plane, we obtain the dilepton exclusion curves in Fig.~\ref{fig:muQmuL}.
We also include the projected dilepton limit with 10 fb$^{-1}$ at ATLAS.

The dijet exclusion limits are identical to those in
the universal bulk mass case if we replace $\mu$ with $\mu_Q$, 
as shown in Fig.~\ref{fig:muL} but in the $(\mu_Q, R^{-1})$ plane, instead. 
We can promote the exclusion curve to 2 dimensional `exclusion surface' as well,
by sweeping it through the $\mu_L$ direction.
Intersecting the resultant surface with the relic abundance surface
yields the dijet exclusion curves in Fig.~\ref{fig:muQmuL}. 
The dilepton limit is more stringent than the dijet limit with relatively large
$\mu_L$. At small $\mu_L$, the dijet limit dominates,
 excluding $\mu_Q$ above 0.7 (0.4) for vanishing $\mu_L L$ in
the case with (without) resonance annihilation. 
In the case of resonant annihilation, we do not show 
the CDF dijet limit in Fig. \ref{fig:muQmuL}(b), 
since it is completely inside the region  excluded by oblique corrections plus relic density.

 If the oblique correction fit C.L. is lowered from 99\%,
 the excluded regions in Fig.~\ref{fig:muQmuL} expand. 
In the no-resonance case at 95\% C.L.,
 an additional excluded `island' region (not shown) appears near
 $\mu_Q L\sim\mu_L L\sim 0.5$ with $R^{-1}\sim 600 \gev$,
corresponding to the portion of the `a' relic density curve near
 $\mu_ R\sim 0.3$ and $R^{-1}\sim 600 \gev$ in Fig.~\ref{fig:muL}
that is excluded by the 95\% C.L. oblique fit contour.
As we lower the fit C.L., this island region grows along the diagonal line 
and eventually merges with the `mainland' exclusion.
In the resonance case, there is no such `island' feature for
any C.L. fit we consider.

Overall, the SUED model with 2 bulk masses is also strongly constrained. 
The yellow-shaded region in Fig. \ref{fig:muQmuL} shows the allowed space, 
which is bounded by electroweak precision constraints
and relic abundance 
at relatively large $\mu_L$, 
and by collider dijet limit at small $\mu_L$. 
In between they are bridged by limits from the dilepton search.
The UED scale $R^{-1}$ is constrained to be between 500-850 GeV without
 $h_2$ resonance annihilation. With resonance, the range is raised to 
 between 650-950 GeV.

\section{Summary and Outlook}
\label{sec:summary}
In general vectorlike fermion-bulk masses could be introduced in models with universal extra dimensions, and 
lead to significant changes in dark matter and collider phenomenology. 
They could shift all KK fermion masses and induce tree-level couplings between even numbered KK gauge bosons and SM fermions. 
They still preserve KK parity, providing a viable dark matter candidate. 
Models with fermion-bulk masses resemble Split SUSY in the sense that partners of SM fermions are heavy. 

We restricted ourselves to the case of positive bulk mass, to keep KK photon as a dark matter. 
In principle, negative bulk mass is also allowed, in which case KK fermions get negative corrections in masses and KK neutrinos will likely be a dark matter candidate. However, direct detection experiments disfavor KK neutrino dark matter  due to 
large elastic scattering cross sections. 

In this paper, we have investigated various constraints including oblique corrections, four Fermi interactions, relic abundance and collider bounds. 
As in the case of MUED, we find a tension between electroweak precision constraints and relic abundance of KK photon for the universal fermion-bulk mass term. 
The remaining parameter space is highly constrained by searches at colliders. 
As a result, we find the allowed region within $\mu R \lesssim$ 0.2-0.3 and $650 \gev \lesssim R^{-1} \lesssim 950 \gev$.
Precise calculation of the relic abundance with 1-loop corrected mass spectrum including all coannihilations and relevant resonances may slightly broaden the 
allowed range for the universal bulk mass term.
However, considering performance of the LHC, the tension still exists and the universal bulk mass is disfavored. 

This tension may be relieved by introducing separate bulk masses for the quark and lepton sectors. 
In this scenario, dijet searches at the LHC together with relic abundance set strong bounds on the bulk parameter in the quark sector, $\mu_Q L \lesssim 0.7$ ($\mu_Q R \lesssim 0.45$), 
while upper limit on the bulk mass in the lepton sector is given by a combination of relic abundance and oblique corrections, $\mu_L L \lesssim 1.7$ ($\mu_L R \lesssim 1.1$).
The LHC is expected to play an important role in constraining the remaining parameter space. 
                                   
Finally we note that a treatment in terms of the oblique $(S,T,U)$ parameters may be insufficient, and 
a global fit to the experimental data is required for a reliable electroweak analysis, 
although our results on oblique corrections contain main features of electroweak constraints.

\acknowledgments
We thank Thomas Flacke for valuable discussion and comments on manuscript. 
GH is supported by the US DOE Grant Number DE-FG02-04ER14308. 
KK is supported in part by the University of Kansas General Research Fund allocation 2301566, 
by the National Science Foundation under Award No. EPS-0903806 and 
matching funds from the State of Kansas through the Kansas Technology Enterprise Corporation. 
SC is supported by Basic Science Research Program through the National Research Foundation of Korea 
funded by the Ministry of Education, Science and Technology (2011-0010294) and (2011-0029758).




\end{document}